\documentclass[12pt,a4paper]{article}
\usepackage[T1]{fontenc}
\usepackage[utf8]{inputenc}
\usepackage{mathptmx,courier}
\usepackage{amsmath,amssymb,graphicx,booktabs,array,tabularx,longtable,url}
\makeatletter
\renewcommand\section{\@startsection{section}{1}{0pt}{3ex plus 1ex minus .2ex}{1.2ex plus .2ex}{\normalfont\large\bfseries\raggedright\hyphenpenalty=10000}}
\renewcommand\subsection{\@startsection{subsection}{2}{0pt}{2.2ex plus .5ex minus .2ex}{.8ex plus .2ex}{\normalfont\normalsize\bfseries\raggedright\hyphenpenalty=10000}}
\makeatother
\newcolumntype{Y}{>{\raggedright\arraybackslash}X}

\newcommand{\href}[2]{#2}
\newcommand{\mmtwo}{\ensuremath{\mathrm{mm}^{2}}}
\newcommand{\mmthree}{\ensuremath{\mathrm{mm}^{3}}}
\newcommand{\legendtitle}[2]{\par\medskip\textbf{Figure #1. #2}\par}
\newcommand{\plate}[3]{\clearpage\begin{center}\includegraphics[width=\linewidth,height=0.78\textheight,keepaspectratio]{#2}\par\medskip\textbf{Figure #1. #3}\end{center}}
\newcommand{\subhead}[1]{\par\medskip\textbf{#1}\par}
\newenvironment{algorithmblock}[1]{\par\medskip\begin{minipage}{\linewidth}\linespread{1.05}\selectfont\small\hrule\smallskip\textbf{#1}\par\smallskip\setlength{\parskip}{3pt}}{\par\smallskip\hrule\end{minipage}\par\medskip}
\newcommand{\algline}[2]{\par\noindent\hspace*{#1em}#2}
\newcommand{\splate}[3]{\clearpage\begin{center}\includegraphics[width=\linewidth,height=0.76\textheight,keepaspectratio]{#2}\par\medskip\textbf{Figure S#1. #3}\end{center}}

\begin{document}
\begin{titlepage}
\linespread{1.0}\selectfont
{\Large\bfseries Reliability assessment and multicenter clinical application of magnetic resonance methods for knee cartilage quantification\par}
\vspace{5mm}
\textbf{Running title:} Knee cartilage quantification

Binbin Yang\textsuperscript{1,*,\ensuremath{\dagger}}, Yongmei Jian\textsuperscript{2,\ensuremath{\dagger}}, Chenglei Liu\textsuperscript{3}, Rui Huang\textsuperscript{4}, Hongda Shao\textsuperscript{5}, Linjun Tong\textsuperscript{6}, Yuanjing Xu\textsuperscript{1}, Jingshu Wu\textsuperscript{1}, Chengzhang He\textsuperscript{1}, Suting Peng\textsuperscript{1}, Ming Xiao\textsuperscript{2}, Yinan Chen\textsuperscript{1,*}, Qi Duan\textsuperscript{1,*}

\textsuperscript{\ensuremath{\dagger}}Binbin Yang and Yongmei Jian contributed equally to this work.

\begingroup\small\setlength{\parskip}{3pt}
\textsuperscript{1}Academy for Clinical Innovation and Translation of Shanghai Co., Ltd. (ACITS), Shanghai 201210, China\par

\textsuperscript{2}Shanghai IxCell Biotechnology Co., LTD, Shanghai 201203, China\par

\textsuperscript{3}Shanghai Ninth People's Hospital, Shanghai Jiao Tong University School of Medicine, Shanghai 200011, China\par

\textsuperscript{4}Nanyang Technological University, Singapore 639798, Singapore\par

\textsuperscript{5}Renji Hospital, Shanghai Jiao Tong University School of Medicine, Shanghai 200127, China\par

\textsuperscript{6}Shanghai Jiading District Central Hospital, Shanghai 201800, China\par
\endgroup

\begin{sloppypar}
\textbf{*Correspondence:} Binbin Yang, \href{mailto:yangbinbin@acits.org.cn}{yangbinbin@acits.org.cn}; Yinan Chen, \href{mailto:chenyinan@acits.org.cn}{chenyinan@acits.org.cn}; Qi Duan, \href{mailto:duanqi@acits.org.cn}{duanqi@acits.org.cn}.
\end{sloppypar}

\textbf{Article type:} Original Article\par
\textbf{Special issue:} Quantitative Imaging in Musculoskeletal Research

\textbf{Trial registration:} Phase II, CTR20210039; phase III, NCT06716281.

\textbf{Ethics approval:} Ethics Committee of Shanghai Ninth People's Hospital, Shanghai Jiao Tong University School of Medicine; phase II, SH9H-2020-C35-2; phase III, SH9H-2023-C13-2.

\textbf{Informed consent and data authorization:} Phase III participants provided written informed consent. Shanghai IxCell Biotechnology Co., LTD authorized the use of de-identified phase II and III images for this study and publication.

\textbf{Funding:} The phase II and phase III clinical trials were sponsored by Shanghai IxCell Biotechnology Co., LTD.

\textbf{Conflicts of interest:} Yongmei Jian and Ming Xiao are employees of Shanghai IxCell Biotechnology Co., LTD, the sponsor of the phase II and phase III clinical trials. The other authors declare no competing interests.

\textbf{Data and software availability:} Individual-level phase II and phase III clinical-trial magnetic resonance imaging (MRI) data, segmentation masks, and derived participant-level data are not publicly available for external distribution. Their use is restricted by clinical-trial confidentiality, data-owner authorization, ethics requirements, and applicable regulatory procedures. Materials required for review by the Center for Drug Evaluation (CDE) of the National Medical Products Administration may be provided to the CDE through the applicable regulatory process. The software implementation is not currently publicly released; future release will depend on regulatory, intellectual-property, and data-governance review and may otherwise remain restricted to CDE review and other authorized use.

\textbf{Permissions:} Not applicable. All figures, tables, and text in this manuscript are original to the present study; no material has been reproduced from another source.

\textbf{Acknowledgments:} We thank the study participants and the clinical-trial teams for their contributions.
\end{titlepage}

\section*{Abstract}
\textbf{Background:} Cartilage volume, thickness, and defect area characterize different aspects of knee cartilage morphology. This study evaluated interreader agreement and longitudinal measurement performance of magnetic resonance-based cartilage quantification methods.

\textbf{Methods:} AI-presegmented masks from 1,189 phase III examinations underwent independent correction by two readers and third-reader adjudication to establish three-structure reference masks. Volume, ray-tracing mean thickness, and defect area defined by a 1.5-mm thickness threshold were calculated. Segmentation and interreader agreement were assessed; the 3D-RBA normal-ray threshold framework was evaluated in 120 three-region digital-phantom experiments from 40 participants. Longitudinal changes were analyzed in 374 participants; alternative methods were compared in a 65-participant subset with progressively increasing volume. Four-visit images from 24 phase II participants were analyzed retrospectively.

\textbf{Results:} AI presegmentation versus adjudicated masks yielded overall Dice of $0.964\pm0.029$. Interreader intraclass correlation coefficients for volume, thickness, and defect area were 0.956, 0.904, and 0.932, respectively; corresponding minimal detectable changes at 95\% confidence (MDC95) were 1,596.9 \mmthree, 0.227 mm, and 147.4 \mmtwo. Among 374 participants, volume changes correlated positively with thickness changes ($\rho=0.431$) and negatively with defect-area changes ($\rho=-0.221$). Mean absolute percentage error for defect-area measurements in the three-region geometric experiments was 5.62\%, with spatial Dice of 0.961. Within-participant correlations in both phase II groups followed the same directions.

\textbf{Conclusions:} The quantification workflow demonstrated good interreader agreement. Geometric errors and longitudinal associations further supported the feasibility of threshold-based defect-area estimation. Volume, thickness, and the extent of involvement provide complementary information for multidimensional assessment of cartilage structure.

\textbf{Keywords:} knee osteoarthritis; magnetic resonance imaging; articular cartilage; artificial intelligence; quantitative imaging; clinical trial.

\textbf{Related trial registrations:} CTR20210039; NCT06716281.

\clearpage
\section{Introduction}
Structural assessment of knee osteoarthritis requires reproducible, interpretable imaging measures. Magnetic resonance imaging (MRI) enables cartilage volume and thickness quantification and is incorporated into structural assessment frameworks for osteoarthritis trials~\cite{r1,r2}. Artificial intelligence (AI) can provide initial segmentations, but complex surfaces, degenerative boundaries, and multicenter acquisition differences still require manual quality control~\cite{r3,r4}.

Volume alone cannot characterize the extent of local damage; thickness and defect area are also influenced by surface correspondence, thresholds, and aggregation rules~\cite{r5,r6,r7,r8}. This study evaluated the reliability of a cartilage quantification framework combining three-dimensional ray-based thickness and defect-area estimation with volume measurement. We assessed interreader agreement, geometric error, longitudinal relationships, and retrospective clinical application within a multicenter human-in-the-loop workflow.

\section{Materials and methods}
\subsection{Study design and ethics}
This study was nested within a randomized, double-blind, placebo-controlled phase III trial of human umbilical cord mesenchymal stem cell injection (IxCell hUC-MSC-O) for knee osteoarthritis across 14 centers (NCT06716281). ACITS served as the independent image-reading center and remained blinded to phase III treatment allocation throughout. Principal eligibility criteria included age 40--80 years, body mass index $\leq30$ kg/m$^2$, and target-knee Kellgren--Lawrence grade II--III disease. Major exclusions included secondary osteoarthritis, substantial deformity, recent knee injections or surgery, and MRI contraindications (Supporting Methods S1.1). Participants received a single intra-articular injection of $5.0\times10^{7}$ cells in 2.0 mL into the target knee or 2.0 mL saline placebo. MRI visits were baseline (V0), week 24 (V6), and week 48 (V8).

All 1,189 phase III examinations underwent independent mask correction by two readers followed by adjudication and were included in presegmentation and interreader agreement analyses. One baseline examination acquired on the incorrect side was retained for these technical agreement analyses, which compared masks derived from the same image, but excluded from target-knee clinical quantification. The clinical quantification dataset therefore comprised 1,188 examinations from 416 participants. Longitudinal analyses included 374 participants with all three visits, of whom 65 met the total-volume criterion V0$<$V6$<$V8. During phase III, historical MRI from the unblinded phase II trial (CTR20210039) was retrospectively annotated using the same workflow. This cohort included 24 participants with four complete visits (18 treated, six controls): screening baseline (V0) and post-dose days 29 (V4), 85 (V5), and 169 (V6). Phase II annotation and adjudication remained blinded; treatment groups were subsequently used for retrospective analysis. Table~\ref{tab:sets} summarizes the analysis sets.

The phase II and III studies followed the Declaration of Helsinki and Good Clinical Practice and were approved by the Ethics Committee of Shanghai Ninth People's Hospital, Shanghai Jiao Tong University School of Medicine (SH9H-2020-C35-2, December 3, 2020; SH9H-2023-C13-2, reapproval July 6, 2023). Participating phase III centers completed required ethics review or filing. All phase III participants provided written informed consent. The sponsor, Shanghai IxCell Biotechnology Co., LTD, authorized use of the de-identified images for this study and publication.

\subsection{Image acquisition and reference standard}
Acquisition requirements included 1.5- or 3.0-T scanners, knee coils with at least four channels, and site-adapted three-dimensional cartilage sequences, with 1.5-mm slice thickness and 0.22--0.35-mm in-plane resolution. Equipment and protocols were kept consistent within participants whenever possible. The independent reading center checked laterality, coverage, motion artifacts, and sequence parameters (Table S1).

Presegmentation used three-dimensional full-resolution self-configuring neural-network (nnU-Net) models~\cite{r9}. V1.0 comprised separate tibiofemoral and patellar models. Tibiofemoral development used 507 OAI-ZIB examinations~\cite{r24}, 100 SKI10 examinations~\cite{r25}, and 119 examinations from the medical big-data training facility operated by the Shanghai Hospital Development Center (SHDC), with vendor-generated automatic labels; patellar development used the latter 119. V2.0 used 428 manually annotated examinations from the SHDC facility: 400 in the development pool (320 training and 80 validation in fold 0) and 28 for separate internal validation. Its single model segmented all three cartilage structures. Both versions used fold 0 for training and model selection; weights were frozen before deployment. V1.0 and V2.0 were deployed for 700 and 489 examinations, respectively. Development and internal-validation data did not overlap with trial deployment data at either participant or examination level (Supporting Methods S1.3).

Model training, presegmentation, and two-reader correction with third-reader adjudication were performed within the SHDC facility~\cite{r10,r11}. Readers underwent standardized training and qualification assessment. Two readers, blinded to participant identity, clinical information, and treatment allocation, independently corrected the same initial mask without access to each other's annotations. They reviewed original images slice by slice, correcting boundaries and connectivity where necessary. An adjudicator, additionally blinded to reader identity, selected the annotation best matching the original images; inadequate annotations were returned for relabeling. Adjudicated masks provided the study reference standard and quantification input for femoral cartilage (FC), tibial cartilage (TC), and patellar cartilage (PC). Calculation rules were consistent across visits, corrections and decisions were logged, and results were reviewed before database lock. AI metrics represent agreement within this human-in-the-loop workflow.

\subsection{Three-dimensional ray-based defect-area estimation}
Regional volume was the cartilage voxel count multiplied by physical voxel volume; total volume was the sum for FC, TC, and PC. Femoral cartilage excluded osteophytes and the intercondylar notch. Marching Cubes extracted the 0.5 isosurface~\cite{r12}, followed by 50 smoothing iterations with a relaxation factor of 0.1. Mesh normals and directions toward the bone centroid distinguished the bone-facing inner and joint-facing outer surfaces. All distances and triangle areas were calculated in physical coordinates. Three-dimensional ray tracing (3D-RT) searched up to 100 mm from each inner-surface vertex along the local normal directed away from bone for the nearest forward outer-surface intersection~\cite{r5}. The distance defined local thickness; absent intersections were assigned zero, and values of 0--10 mm were retained. The nearest distance was recorded once per returned intersection. Regional means, including zero values, were rounded to two decimal places; overall thickness was their unweighted mean.

Three-dimensional ray-based area (3D-RBA) defined involvement by residual cartilage thickness. International Cartilage Repair Society (ICRS) grade III lesions exceed half the cartilage depth without penetrating subchondral bone, whereas grade IV lesions involve subchondral bone~\cite{r13}. Mean healthy tibiofemoral cartilage thickness is approximately 2.75 mm, or roughly 3 mm, with regional and population variation~\cite{r14,r15,r16}. Against this approximate reference, residual thickness below 1.5 mm corresponds to loss of more than half the layer, motivating a continuous measure of the extent of marked thinning. Accordingly, 1.5 mm was prespecified as an operational threshold; it was neither optimized against study results nor intended to replace ICRS grading.

For an inner-surface point $p$, the first valid outer-surface intersection $q$ along its local normal defined $t(p)=\lVert q-p\rVert$. Outer-surface triangles associated with intersections satisfying $0<t(p)<1.5$ mm formed $A_{\mathrm{thin}}$. Where no valid intersection existed, neighboring inner-surface triangles were included according to prespecified boundary-compensation rules, forming $A_{\mathrm{zero}}$. After deduplication, defect area was calculated from triangle vertices in physical coordinates:
\begin{equation}
\mathrm{3D\mbox{-}RBA}=\sum_{f\in A_{\mathrm{thin}}\cup A_{\mathrm{zero}}}
\frac{1}{2}\left\lVert (p_2-p_1)\times(p_3-p_1)\right\rVert.
\end{equation}
Total defect area was the sum across FC, TC, and PC, in \mmtwo.

This definition encompasses focal or diffuse thinning with residual cartilage below the threshold and candidate full-thickness defects without measurable thickness. The latter are estimated through boundary compensation; natural cartilage margins are processed under the same rules. Surface construction and defect-area pseudocode are detailed in Supporting Methods S2.

The alternative three-dimensional patch-mapped area (3D-PMA) method drew on cartilage-coverage and denuded-bone research~\cite{r2,r17,r18,r19} to estimate the bone-reference surface not covered by cartilage, assessing sensitivity to defect-area definition. Within the cartilage-adjacent bone region, triangle fractions exceeding an adaptive bone--cartilage contact-distance threshold contributed to the defect-area estimate. Thickness comparators were three-dimensional nearest-neighbor (3D-NN), normal-matching (3D-MN), and signed-distance-transform (3D-SDT) methods~\cite{r5,r20}. 3D-NN measured the shortest outer-to-inner surface distance. 3D-MN selected an inner-surface neighbor with minimum deviation from the local bone-referenced inward direction and used its projected distance. 3D-SDT approximated local thickness as twice the physical distance from an interior cartilage voxel to the boundary. All three retained positive finite values and used 1\% trimming at each tail when at least 50 values were available. 3D-NN and 3D-MN were aggregated by pretrimming valid-point counts, and 3D-SDT by cartilage volume. Thickness algorithms and aggregation rules are provided in Supporting Methods S2.1 and alternative defect-area estimation in S2.3.

\subsection{Methodological evaluation and statistical analysis}
AI agreement was evaluated using Dice and intersection over union (IoU) for FC, TC, PC, and their union. Six-connected surface voxels were identified, and bidirectional nearest-surface distances were calculated in physical coordinates. The pooled 95th percentile defined the 95th-percentile Hausdorff distance (HD95); the average of the two directional means defined average surface distance (ASD). Interreader volume, 3D-RT, and 3D-RBA were assessed using the two-way random-effects, absolute-agreement, single-measurement intraclass correlation coefficient, ICC(2,1)~\cite{r21}. Each reader's independently corrected mask was processed with the same quantification methods for all 1,189 examinations. Absolute agreement was further assessed using repeated-measures Bland--Altman analysis, with paired differences defined as Reader A minus Reader B~\cite{r26}. Participant and residual components of the paired-difference variance were estimated using a random-intercept variance decomposition; 95\% limits of agreement (LoA) were calculated as the mean bias $\pm1.96$ times the total difference standard deviation. The standard error of measurement (SEM) was the total difference standard deviation divided by $\sqrt{2}$, and the interreader minimal detectable change at 95\% confidence (MDC95) was $1.96\sqrt{2}\times\mathrm{SEM}$~\cite{r27}. Participant-level bootstrap resampling (2,000 iterations), retaining all examinations from each sampled participant, estimated 95\% confidence intervals (CIs). Proportional bias was assessed by regression of paired differences on paired means with participant-clustered standard errors.

Forty participants were selected by simple random sampling without replacement from the 416-participant phase III imaging cohort. Each participant contributed a baseline femoral (FC), tibial (TC), and patellar (PC) cartilage mask, yielding 120 three-region geometric experiments. Cartilage was first expanded to a 3-mm baseline; one continuous internal region of approximately 100 \mmtwo\ per cartilage structure was then thinned to 1-mm residual thickness, away from natural margins. Isotropic voxel size was 0.5 mm. The reference defect area comprised outer-surface triangles within the predefined design footprint whose independently sampled Euclidean distance to bone was $<1.5$ mm. The 3D-RBA normal-ray threshold framework was tested by ray-based classification and summation of selected triangle areas. Post-thinning total defect area was the geometric estimate. Errors and spatial overlap were summarized across regions, with participant-level resampling for confidence intervals (CIs; Supporting Methods S3).

Longitudinal analyses in 374 participants were not selected by change direction; Spearman correlations were calculated between final-minus-baseline changes. The 65-participant increasing-volume subset supported alternative-method comparisons and complete individual trajectories. All methods used the same adjudicated masks; no additional selection was based on thickness or defect-area changes. Percentage changes were differences between final and baseline visit means divided by the baseline mean. Phase II means and standard deviations were summarized by group and visit. Pearson correlations were calculated after subtracting each participant's four-visit mean. Correlation CIs used 5,000 participant-level bootstrap iterations, retaining each participant's observations together. Phase II analyses did not adjust for visit effects or test between-group treatment effects. Statistics and plotting used Python 3.12 with NumPy, pandas, SciPy, and Matplotlib.

\section{Results}
\subsection{Segmentation and interreader agreement}
Against adjudicated masks, AI presegmentation across 1,189 examinations yielded overall Dice, IoU, HD95, and ASD of $0.964\pm0.029$, $0.932\pm0.052$, $0.533\pm0.597$ mm, and $0.098\pm0.109$ mm, respectively; 79.2\% had overall Dice $\geq0.95$ (Figure~\ref{fig:agreement}; Table S2). Patellar cartilage showed relatively lower overlap. Distributions by visit and deployed version are provided in Figures S1--S4. Three-dimensional reconstructions of adjudicated masks depict anatomical contours and surface variations, providing spatial context for quantification (Figure~\ref{fig:reconstruction}).

Overall interreader ICCs were 0.956 (95\% CI, 0.947--0.963) for volume, 0.904 (0.888--0.918) for 3D-RT, and 0.932 (0.917--0.946) for 3D-RBA; regional agreement was relatively lower for patellar cartilage (Table~\ref{tab:icc}). Repeated-measures Bland--Altman biases (Reader A minus Reader B) were 367.7 \mmthree\ for volume, 0.030 mm for 3D-RT, and 6.9 \mmtwo\ for 3D-RBA, with 95\% LoA of $-1,229.1$ to 1,964.6 \mmthree, $-0.197$ to 0.257 mm, and $-140.5$ to 154.3 \mmtwo, respectively. Corresponding interreader MDC95 values were 1,596.9 \mmthree\ (95\% CI, 1,478.9--1,715.9), 0.227 mm (0.213--0.239), and 147.4 \mmtwo\ (132.6--163.8). Small proportional trends were detected, explaining 1.9\%--4.6\% of difference variance (Table S6; Figure S7). Adjudicated cartilage masks matched only reader 1 in 1,075 examinations, only reader 2 in 107, and both readers in seven.

\subsection{Defect-area errors in geometric models}
Across 120 experiments in 40 participants, mean reference and measured post-thinning defect areas were 86.793 and 83.675 \mmtwo, respectively, with a bias of $-3.119$ \mmtwo. Mean absolute error was 4.627 \mmtwo\ (95\% CI, 4.180--5.113), mean absolute percentage error was 5.616\% (4.946\%--6.457\%), concordance correlation coefficient was 0.863, and local spatial Dice was 0.961. Absolute percentage errors were $\leq10\%$ in 113/120 experiments and $\leq15\%$ in 119/120. Regional mean absolute percentage errors were 6.461\% for FC, 3.504\% for TC, and 6.884\% for PC (Figure~\ref{fig:geometry}; Table S3).

\subsection{Longitudinal relationships and alternative methods}
Among 374 complete cases, volume changes correlated positively with 3D-RT changes ($\rho=0.431$) and negatively with 3D-RBA changes ($\rho=-0.221$). Thickness and defect-area changes were also negatively correlated ($\rho=-0.314$); all $P<0.001$ (Table~\ref{tab:correlation}).

In the 65-participant increasing-volume subset, final mean volume was 7.28\% above baseline, and mean thickness increased across visits with all four methods; 3D-RT increased from 2.189 to 2.264 mm. Mean 3D-RBA was 987.922, 991.291, and 938.334 \mmtwo\ at V0, V6, and V8, respectively, remaining close to baseline at V6 and decreasing by 5.02\% at V8; 3D-PMA decreased by 1.59\% at V8 (Figure~\ref{fig:longitudinal}A,B). Changes in 3D-RT correlated positively with comparator thickness changes ($\rho=0.318$--0.458), whereas changes in the two defect-area estimates showed no clear correlation ($\rho=0.027$). Volume changes correlated negatively with 3D-RBA ($\rho=-0.356$) but not clearly with 3D-PMA ($\rho=0.007$; Table~\ref{tab:correlation}).

All 195 examinations in this subset yielded complete measurements across 585 anatomical regions. Complete paired trajectories for total volume, 3D-RT, and 3D-RBA are presented together in Figure~\ref{fig:longitudinal}C. Relative to baseline, final 3D-RT was higher in 46 participants (70.8\%) and final 3D-RBA was lower in 44 (67.7\%). Volume and thickness changes correlated positively ($\rho=0.654$), and defect-area and thickness changes correlated negatively ($\rho=-0.405$). Group distributions and individual trajectories jointly illustrated complementary structural relationships.

\subsection{Retrospective phase II application}
Among 24 participants with 96 four-visit examinations, final mean volume changed by $+0.17\%$ and $-0.10\%$ in the treatment and control groups, respectively; corresponding changes were $-0.97\%$ and $+4.18\%$ for 3D-RBA and $-2.58\%$ and $-1.14\%$ for 3D-RT (Figure~\ref{fig:phaseii}; Table S4). Metrics exhibited distinct longitudinal patterns rather than synchronous monotonic group-mean changes. Within participants, volume correlated positively with thickness and negatively with defect area in the treatment group ($r=0.529$, 95\% CI, 0.322--0.707; $r=-0.429$, $-0.612$ to $-0.253$) and the control group ($r=0.762$, 0.689--0.868; $r=-0.334$, $-0.537$ to $-0.131$). The 95\% CIs for defect-area--thickness correlations crossed zero in both groups (Table S5).

\section{Discussion}
This study evaluated cartilage quantification reliability using volume, 3D-RT thickness, and 3D-RBA defect area. Overall interreader ICCs exceeded 0.90 for all three metrics, including 0.932 for 3D-RBA, supporting measurement agreement despite differences in manual correction. Bland--Altman analysis complemented these relative-reliability estimates by quantifying systematic and random interreader differences in the original units. Mean biases represented 2.48\%, 1.37\%, and 0.71\% of the paired means for volume, 3D-RT, and 3D-RBA, respectively, while the MDC95 values provide workflow-specific thresholds for changes exceeding interreader annotation variability.

3D-RBA translates ICRS lesion-depth concepts and healthy-cartilage thickness information into continuous defect-area estimation under a fixed threshold~\cite{r13,r14,r15,r16}. It combines markedly thinned but still covered regions with candidate defects estimated by boundary compensation, complementing total volume and mean thickness by describing the extent of involvement. The geometric experiments tested the thickness-threshold and triangle-area calculation under controlled continuous thinning. Their error estimates address this prescribed geometry, while the 374-participant analysis examines whether clinical changes follow the expected cross-metric relationships without selecting their direction. Together with interreader agreement, these complementary evaluations support the feasibility and interpretability of 3D-RBA for longitudinal structural assessment.

The increasing-volume subset supported comparisons across methods and individual trajectories. Mean thickness increased with all four methods; final mean 3D-RBA decreased, and its changes correlated negatively with volume, consistent with the full cohort. Whereas 3D-PMA primarily estimates uncovered bone surface, 3D-RBA additionally includes markedly thinned cartilage-covered regions; responses to local thickness changes can therefore differ. The association between 3D-RBA and volume changes supports its use in characterizing marked thinning.

Retrospective phase II analysis applied the same annotation and quantification workflow to another clinical imaging cohort, examining relationships across visits and treatment contexts while maintaining blinded phase III independent reading. Positive within-participant volume--thickness and negative volume--defect-area correlations in both groups provide complementary evidence for clinical application. The metrics need not change synchronously: volume sums cartilage tissue, mean thickness gives equal weight to three regions, and 3D-RBA reflects local threshold-defined involvement.

Medial/lateral anatomical stratification can further characterize spatial heterogeneity. An atlas-assisted partitioning model defined a separating plane while preserving adjudicated cartilage contours, providing consistent spatial rules for regional analysis. This extension underwent physician visual review; the full workflow is presented in Supporting Discussion S6 as a basis for subsequent regional measurement and boundary-repeatability studies. Reliability evaluation in this article remains based on FC, TC, and PC.

The scope of the findings should be considered in relation to the evaluation design. Both readers started from the same AI mask, so agreement evidence pertains to this human-in-the-loop workflow. Applicability of the fixed threshold across populations and devices, and scan--rescan stability, require further evaluation. Longitudinal associations should be interpreted in light of shared masks and cohort selection; the small phase II analysis provides exploratory methodological evidence.

\section{Conclusions}
Volume, 3D-RT, and 3D-RBA in this cartilage MRI quantification workflow demonstrated good interreader agreement. Controlled geometric errors and longitudinal cross-metric associations further support the feasibility of 3D-RBA, providing a methodological basis for assessing cartilage structural change in terms of total amount, thickness, and extent of involvement.

\section*{List of abbreviations}
AI, artificial intelligence; MRI, magnetic resonance imaging; FC, femoral cartilage; TC, tibial cartilage; PC, patellar cartilage; 3D-RT, three-dimensional ray tracing; 3D-RBA, three-dimensional ray-based area; 3D-PMA, three-dimensional patch-mapped area; 3D-NN, three-dimensional nearest neighbor; 3D-MN, three-dimensional normal matching; 3D-SDT, three-dimensional signed distance transform; ICRS, International Cartilage Repair Society; ICC, intraclass correlation coefficient; SEM, standard error of measurement; MDC95, minimal detectable change at 95\% confidence; LoA, limits of agreement; IoU, intersection over union; HD95, 95th-percentile Hausdorff distance; ASD, average surface distance; CI, confidence interval.

\section*{Author contributions and acknowledgments}
\textbf{Author contributions:} \textbf{Binbin Yang:} Conceptualization; methodology; software; validation; formal analysis; visualization; supervision; writing--original draft. He led the technical work, trained the V1.0 model, designed and implemented the volume, thickness, and defect-area methods, designed the experiments, analyzed the results, and prepared the figures and manuscript. \textbf{Yongmei Jian:} Resources; methodology. She provided relevant standards and expert-consensus guidance for the two-reader correction and adjudication workflow. \textbf{Chenglei Liu:} Conceptualization; investigation; validation. He served as a clinical reader in the MRI-based cartilage-mask annotation and review workflow. He also proposed estimating cartilage defect area using the 1.5-mm residual-thickness threshold. \textbf{Rui Huang:} Investigation; data curation; project administration. He performed AI presegmentation of de-identified images, uploaded masks, created and managed annotation tasks, and supported the two-reader and adjudication workflow within the SHDC facility. \textbf{Hongda Shao:} Investigation; validation. He served as a clinical reader in the MRI-based cartilage-mask annotation and review workflow. \textbf{Linjun Tong:} Investigation; validation. He served as a clinical reader in the MRI-based cartilage-mask annotation and review workflow. \textbf{Yuanjing Xu:} Project administration; coordination of the independent image-reading center, readers, and sponsor. \textbf{Jingshu Wu:} Formal analysis; validation. She independently confirmed the statistical results. \textbf{Chengzhang He:} Conceptualization; clinical interpretation of 3D-RBA using ICRS lesion-depth concepts. \textbf{Suting Peng:} Software. She trained the V2.0 AI presegmentation model. \textbf{Ming Xiao:} Resources; project administration. He coordinated and advanced the clinical-trial project on behalf of the sponsor. \textbf{Yinan Chen:} Writing--review and editing, including contextual and English-language revision. \textbf{Qi Duan:} Resources; supervision.

\clearpage
\begingroup\linespread{1.0}\selectfont\small

\par\endgroup

\clearpage
\begingroup\linespread{1.0}\selectfont\small
\refstepcounter{table}\label{tab:sets}
\textbf{Table \thetable. Imaging analysis sets and their purposes}
\par\smallskip
\begin{tabularx}{\textwidth}{@{}>{\raggedright\arraybackslash}p{43mm}>{\raggedright\arraybackslash}p{34mm}Y@{}}
\toprule
Dataset & Participants / examinations & Purpose \\
\midrule
Phase III presegmentation and interreader agreement & 1,189 examinations & Dice, IoU, HD95, and ASD for AI masks versus adjudicated reference masks; interreader ICCs, Bland--Altman LoA, and MDC95 for volume, 3D-RT, and 3D-RBA \\
Phase III final clinical quantification database & 416 participants; 1,188 examinations & Cartilage volume, 3D-RT mean thickness, and 3D-RBA defect area \\
Phase III three principal visits & V0: 402; V6: 393; V8: 392 examinations & Baseline, week 24, and week 48 \\
Phase III nonprincipal visit & V1: 1 examination & Included in the final database, but not in principal-visit figures \\
Phase III complete three-visit cases & 374 participants & Eligible data for paired longitudinal analysis \\
Phase III alternative defect-area and thickness comparisons & 65 participants; 195 examinations & Subset selected by total volume V0$<$V6$<$V8 \\
Phase II retrospective complete four-visit cohort & 24 participants; 96 examinations & 18 treated, six controls; longitudinal distributions and within-participant correlations at V0, V4, V5, and V6 \\
\bottomrule
\end{tabularx}
\par\smallskip\footnotesize
Phase III V0, V6, and V8 denote baseline, week 24, and week 48; phase II V0, V4, V5, and V6 denote screening baseline and post-dose days 29, 85, and 169. One baseline examination acquired on the incorrect side was retained in the 1,189-examination technical agreement set but excluded from the 1,188-examination clinical quantification dataset. The nonprincipal V1 record was excluded from three-visit longitudinal analysis.
\par\endgroup

\clearpage
\begingroup\linespread{1.0}\selectfont\small
\refstepcounter{table}\label{tab:icc}
\textbf{Table \thetable. Interreader agreement for cartilage quantification}
\par\smallskip
\begin{tabularx}{\textwidth}{@{}Yccc@{}}
\toprule
Region & Volume ICC & 3D-RT ICC & 3D-RBA ICC \\
\midrule
Overall & 0.956 (0.947--0.963) & 0.904 (0.888--0.918) & 0.932 (0.917--0.946) \\
Femoral cartilage & 0.965 (0.956--0.973) & 0.943 (0.924--0.958) & 0.924 (0.908--0.939) \\
Tibial cartilage & 0.949 (0.937--0.959) & 0.961 (0.951--0.969) & 0.912 (0.884--0.935) \\
Patellar cartilage & 0.858 (0.835--0.878) & 0.846 (0.820--0.868) & 0.886 (0.853--0.911) \\
\bottomrule
\end{tabularx}
\par\smallskip\footnotesize
Interreader agreement was evaluated in all 1,189 examinations, including one incorrect-side baseline examination retained for paired comparison on the same image but excluded from the 1,188-examination clinical quantification dataset. Values are ICCs (95\% CIs). A two-way random-effects, absolute-agreement, single-measurement model, ICC(2,1), was used with 2,000 participant-level bootstrap iterations. Overall volume and defect area are sums across the three regions; overall thickness is the unweighted mean of the regional means.
\par\endgroup

\clearpage
\begingroup\linespread{1.0}\selectfont\small
\refstepcounter{table}\label{tab:correlation}
\textbf{Table \thetable. Correlations between longitudinal changes in the complete phase III cohort and increasing-volume subset}
\par\smallskip
\begin{tabularx}{\textwidth}{@{}>{\raggedright\arraybackslash}p{44mm}Ycc@{}}
\toprule
Analysis set & Paired V8$-$V0 changes & Spearman $\rho$ & 95\% CI \\
\midrule
Complete, $n=374$ & Volume and 3D-RT & 0.431 & 0.334 to 0.520 \\
Complete, $n=374$ & Volume and 3D-RBA & $-0.221$ & $-0.325$ to $-0.112$ \\
Complete, $n=374$ & 3D-RT and 3D-RBA & $-0.314$ & $-0.417$ to $-0.206$ \\
\midrule
Increasing volume, $n=65$ & 3D-RBA and 3D-PMA & 0.027 & $-0.236$ to 0.282 \\
Increasing volume, $n=65$ & 3D-RT and 3D-NN & 0.389 & 0.121 to 0.633 \\
Increasing volume, $n=65$ & 3D-RT and 3D-MN & 0.458 & 0.210 to 0.674 \\
Increasing volume, $n=65$ & 3D-RT and 3D-SDT & 0.318 & 0.048 to 0.558 \\
Increasing volume, $n=65$ & Volume and 3D-RBA & $-0.356$ & $-0.532$ to $-0.130$ \\
Increasing volume, $n=65$ & Volume and 3D-RT & 0.654 & 0.472 to 0.775 \\
Increasing volume, $n=65$ & 3D-RBA and 3D-RT & $-0.405$ & $-0.619$ to $-0.155$ \\
Increasing volume, $n=65$ & Volume and 3D-PMA & 0.007 & $-0.243$ to 0.267 \\
\bottomrule
\end{tabularx}
\par\smallskip\footnotesize
Each participant contributed one pair of final-minus-baseline changes. CIs used 5,000 participant-level bootstrap iterations. The 374-participant analysis was not selected by change direction; all three correlations had $P<0.001$. The 65-participant subset met total volume V0$<$V6$<$V8; $P=0.004$ for volume versus 3D-RBA and $P=0.956$ for volume versus 3D-PMA. Correlation does not establish absolute measurement agreement or treatment effects.
\par\endgroup

\clearpage
\section*{Figure legends}
\begingroup\linespread{1.0}\selectfont\footnotesize
\legendtitle{1}{Structure-specific agreement between AI presegmentation and adjudicated reference masks.}
All 1,189 examinations were included. A--D show Dice, IoU, HD95, and ASD, respectively. The horizontal axis denotes tibial, femoral, patellar, and overall cartilage; overall cartilage is the union of the three structures. Boxes span the 25th--75th percentiles, horizontal lines indicate medians, whiskers extend to the most extreme values within 1.5 interquartile ranges of the box boundaries, yellow triangles denote means, and points denote outliers. Axes focus on the principal distributions; outliers outside the displayed ranges remain included in all statistics.

\legendtitle{2}{Three-dimensional reconstructions of adjudicated cartilage masks from four participants.}
Reconstructed femoral, tibial, and patellar cartilage surfaces show differences in contours, continuity, and surface morphology. All images derive from masks obtained through two-reader correction and third-reader adjudication, providing spatial context for subsequent volume, thickness, and defect-area quantification.

\legendtitle{3}{Digital-phantom validation of the 3D-RBA normal-ray threshold framework.}
Forty randomly selected participants each contributed baseline FC, TC, and PC masks, yielding 120 controlled artificial-thinning experiments. A compares independently defined reference defect areas within the design footprint with measured post-thinning total defect areas; the dashed line is identity. B shows absolute defect-area errors, with the dashed line denoting the overall mean absolute error. C summarizes absolute percentage errors across the three cartilage regions; boxes span the 25th--75th percentiles, green lines indicate medians, yellow triangles indicate means, and points indicate outliers beyond 1.5 interquartile ranges. D shows mean area-weighted precision, recall, Dice, and IoU within the local evaluation region. FC, femoral cartilage; TC, tibial cartilage; PC, patellar cartilage.

\legendtitle{4}{Metric distributions and individual longitudinal trajectories in the 65-participant increasing-volume subset.}
All panels use the same 65 participants with total volume V0$<$V6$<$V8, contributing 195 examinations. V0, V6, and V8 denote baseline, week 24, and week 48. A shows visit distributions of total volume, 3D-RBA, and 3D-PMA to examine longitudinal responses under different defect-area definitions. B shows 3D-RT, 3D-NN, 3D-MN, and 3D-SDT in the upper-left, upper-right, lower-left, and lower-right plots, respectively, assessing sensitivity to thickness definition. C combines all individual trajectories for total volume (left), 3D-RT (middle), and 3D-RBA (right), extending group distributions to paired observations within participants. Volume and defect area are summed across FC, TC, and PC. 3D-RT is the unweighted regional mean; 3D-NN and 3D-MN are weighted by valid-point counts, and 3D-SDT by volume. In A and B, boxes span the 25th--75th percentiles, green lines indicate medians, whiskers extend to the most extreme values within 1.5 interquartile ranges, red lines and triangles connect visit means, and points denote outliers. Across the three plots in C, each participant retains the same color and marker, with a shared legend; black dashed lines and open triangles indicate means. Vertical axes cover all observations.

\legendtitle{5}{Quantitative metric distributions in the complete four-visit phase II cohort.}
A, total volume; B, 3D-RBA defect area; C, 3D-RT mean thickness. These panels evaluate longitudinal performance of the same quantification workflow in historical clinical images. Every visit includes 18 treated participants and six controls, represented by blue and orange boxplots, respectively. Boxes span the 25th--75th percentiles, green lines indicate medians, and whiskers extend to the most extreme observations within 1.5 interquartile ranges of the box edges. Blue circles and orange squares show all individual values with horizontal jitter. Yellow triangles indicate group means, connected across visits by lines in the corresponding group color. Numeric rows below each panel give the mean $\pm$ standard deviation (SD) by group and visit, as also reported in Table S4. Vertical ranges include all observations. V0 denotes screening baseline; V4, V5, and V6 denote post-dose days 29, 85, and 169. Visits are equally spaced categorical positions. Volume and defect area are sums across the three regions; thickness is the unweighted mean of regional means. Groups are used to describe metric relationships in this retrospective phase II cohort.
\par\endgroup

\refstepcounter{figure}\label{fig:agreement}
\plate{1}{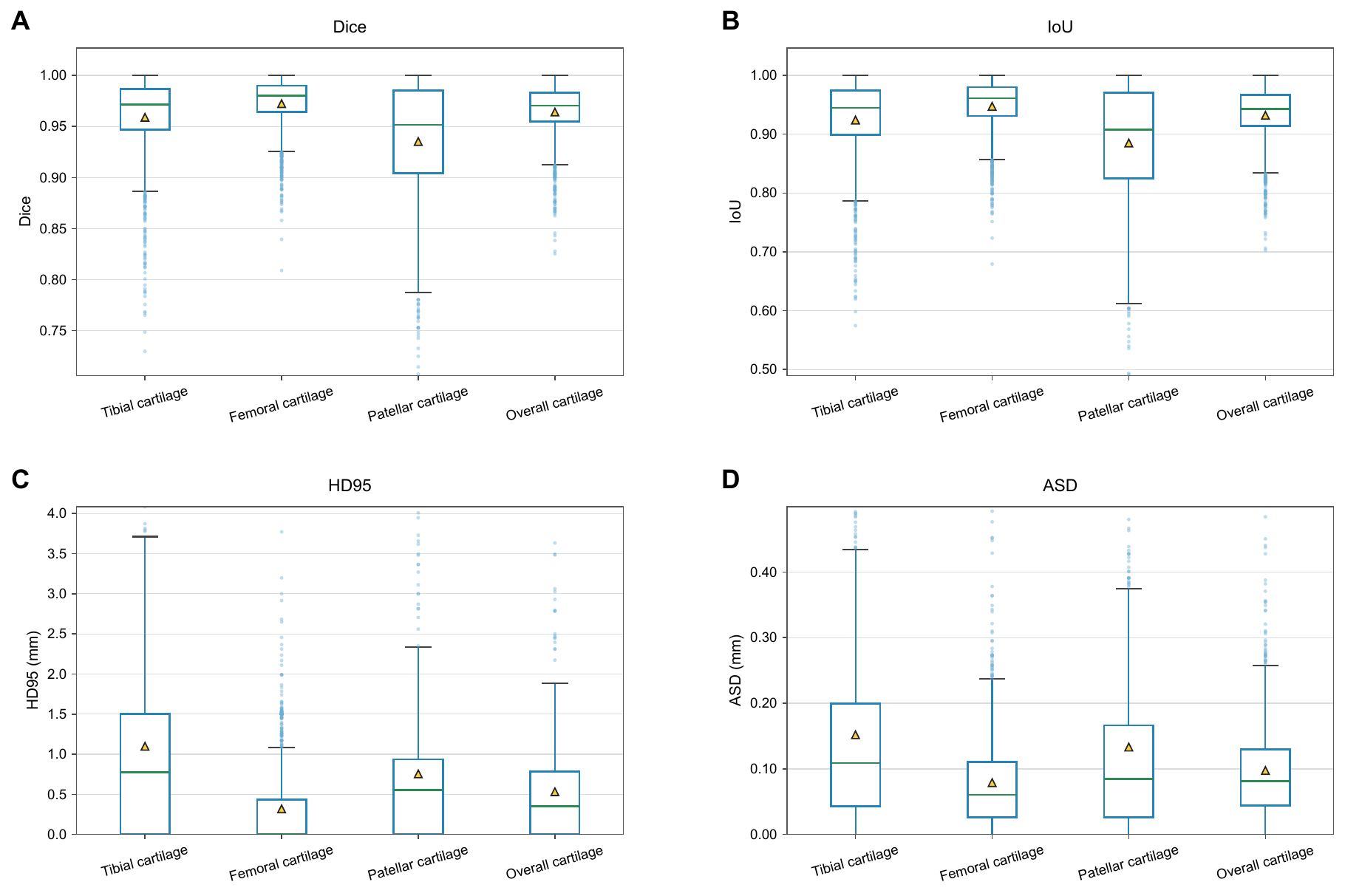}{Structure-specific segmentation agreement}
\refstepcounter{figure}\label{fig:reconstruction}
\plate{2}{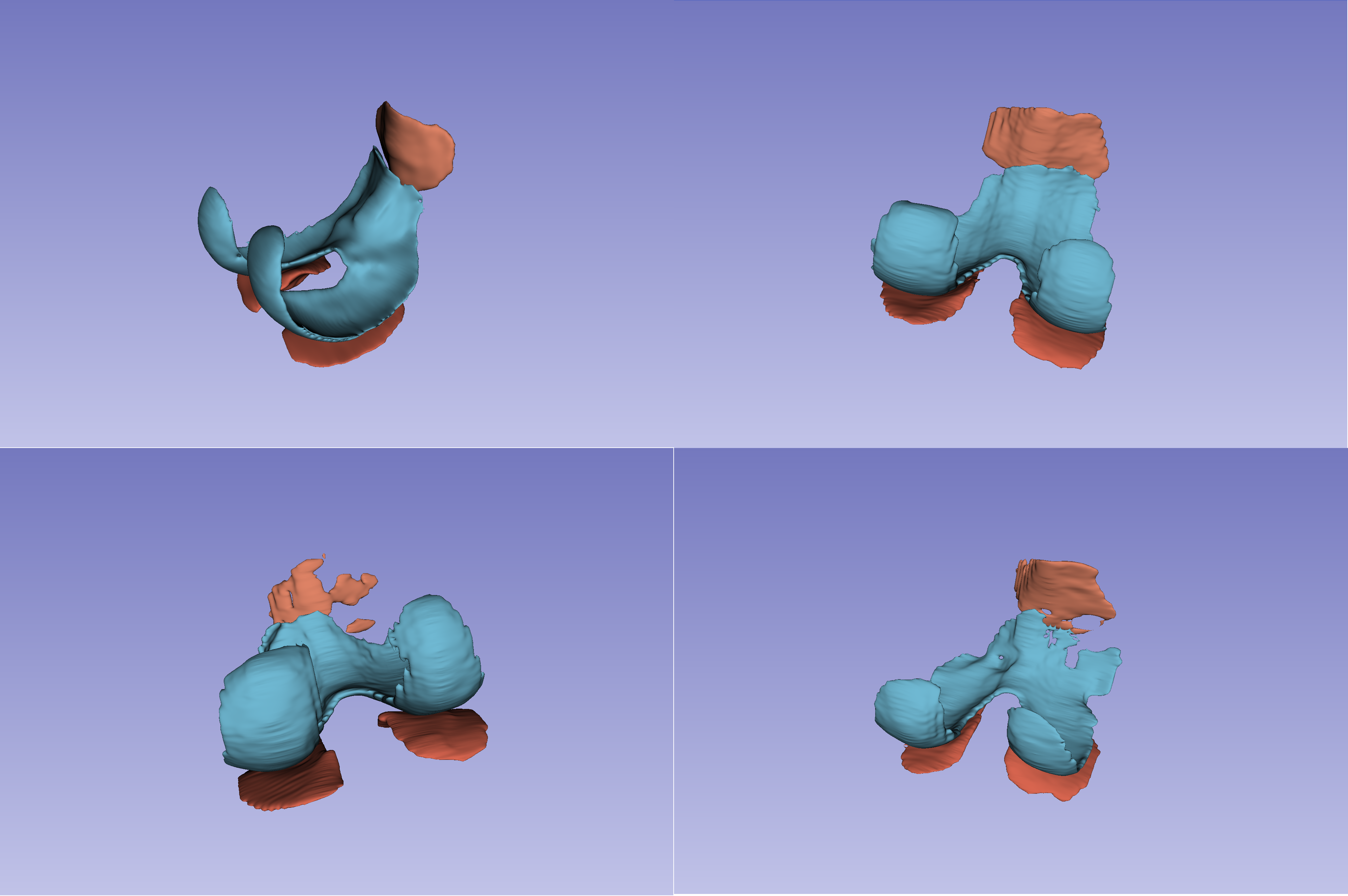}{Three-dimensional cartilage reconstructions}
\refstepcounter{figure}\label{fig:geometry}
\plate{3}{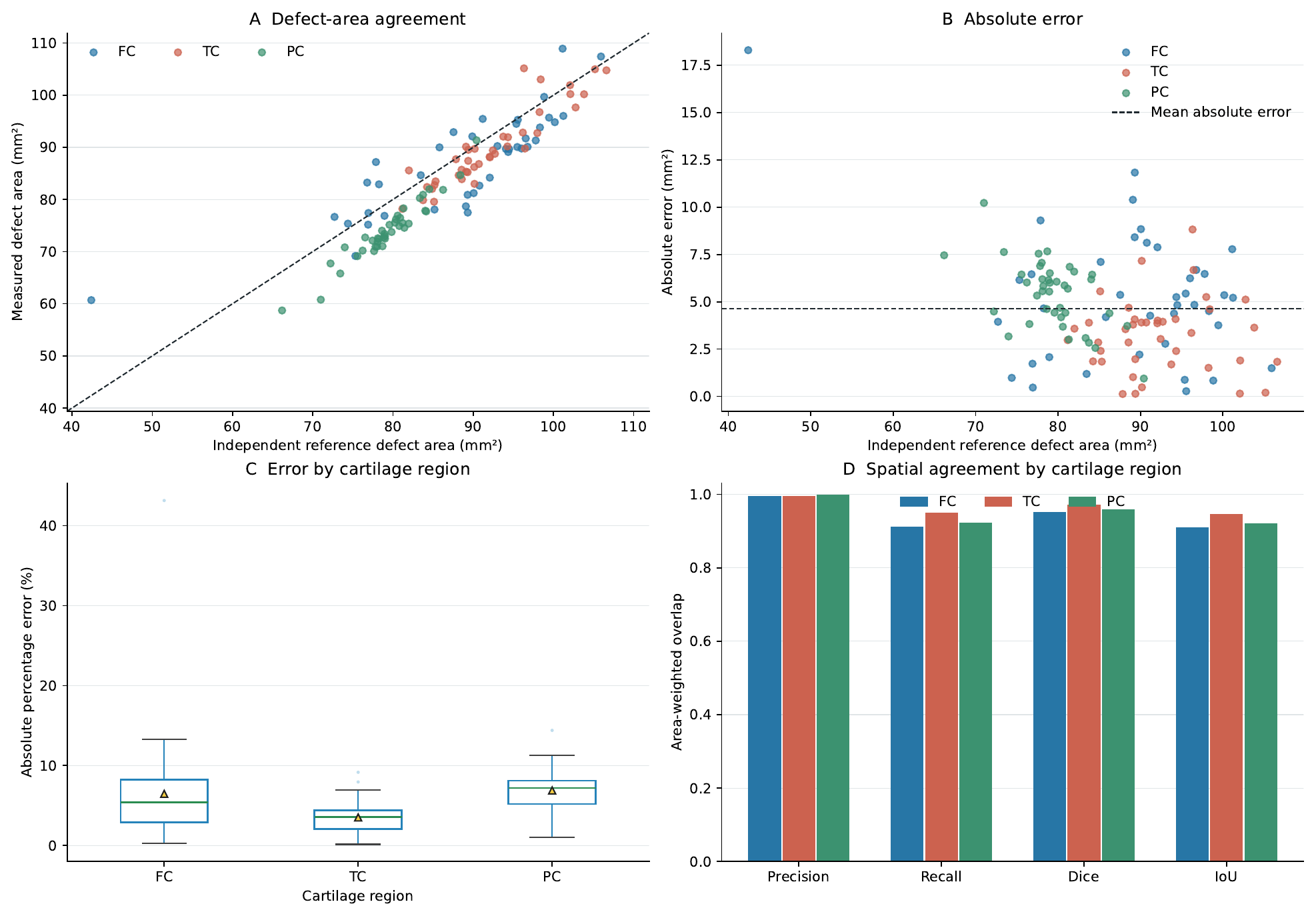}{Controlled geometric evaluation}

\clearpage\refstepcounter{figure}\label{fig:longitudinal}
\begin{center}
\includegraphics[width=\linewidth]{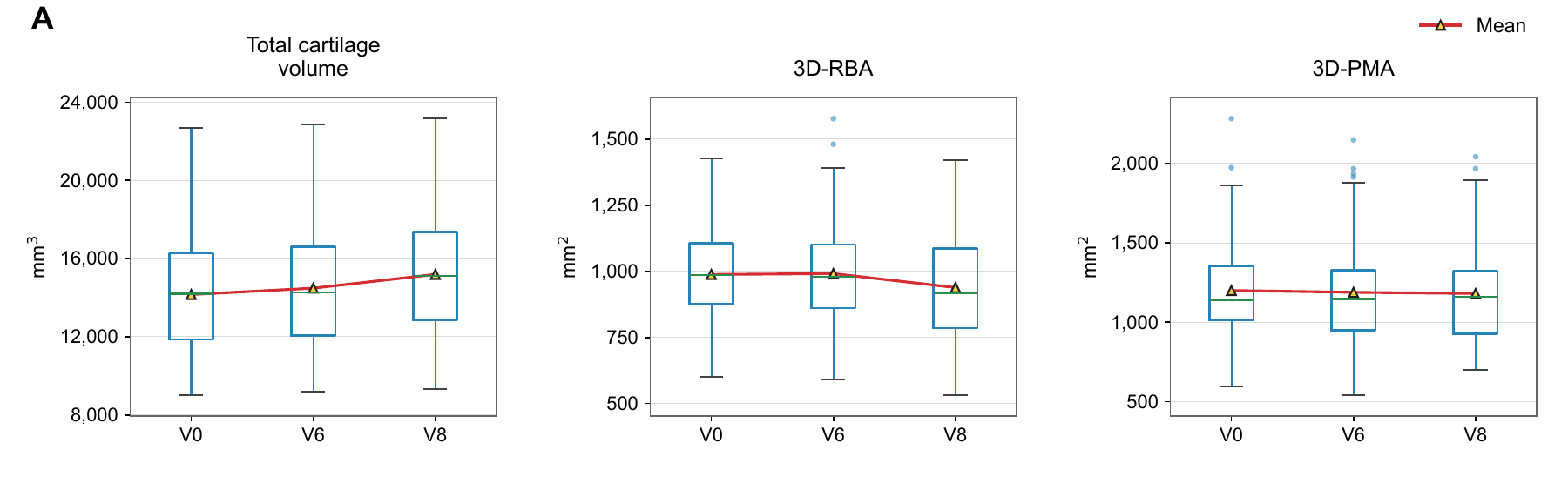}\par\vspace{6mm}
\includegraphics[width=\linewidth]{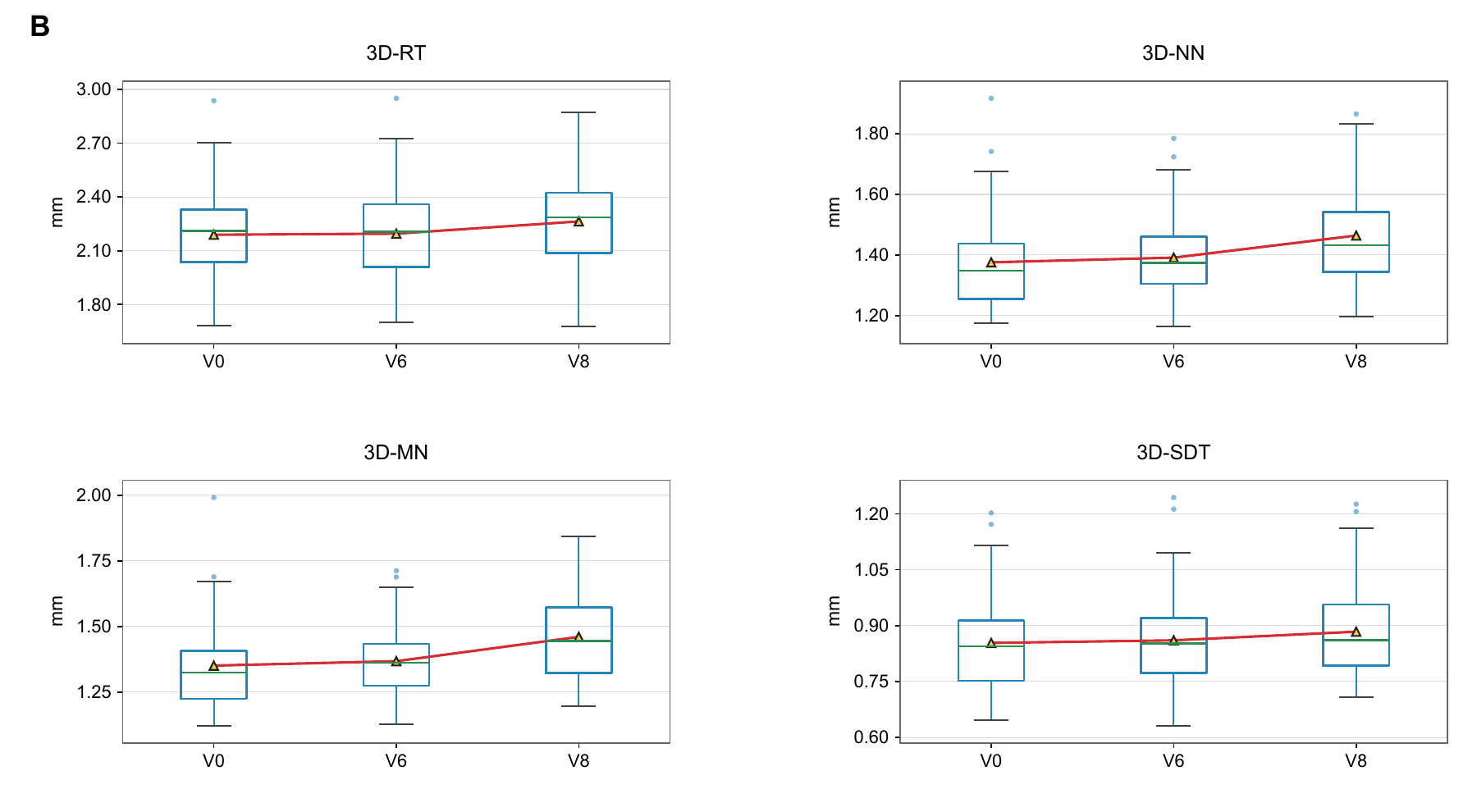}\par\medskip
\textbf{Figure 4. Longitudinal metric distributions (A and B)}
\end{center}
\plate{4 (continued)}{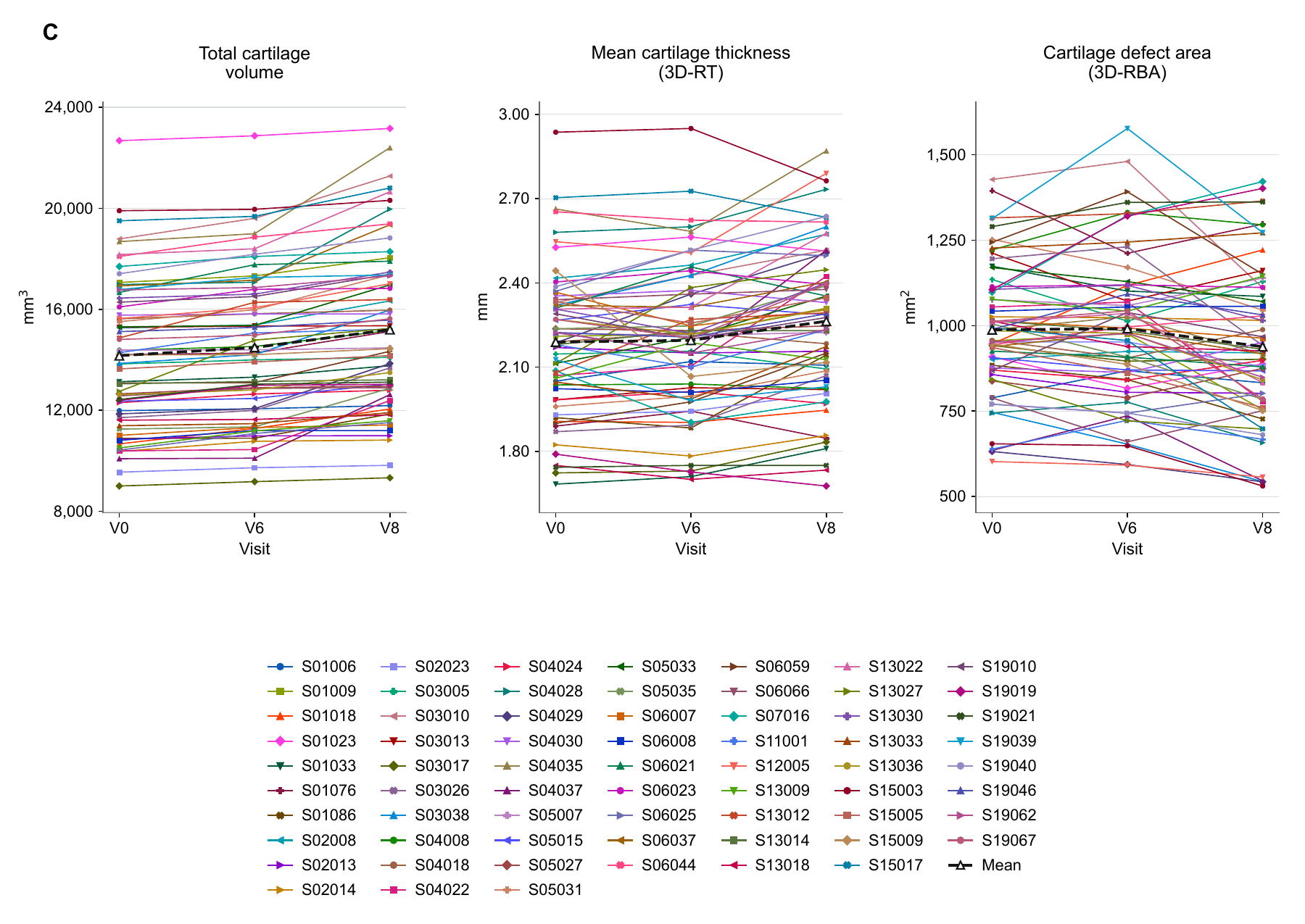}{Individual trajectories (C)}

\clearpage\refstepcounter{figure}\label{fig:phaseii}
\begin{center}
\includegraphics[width=0.94\linewidth]{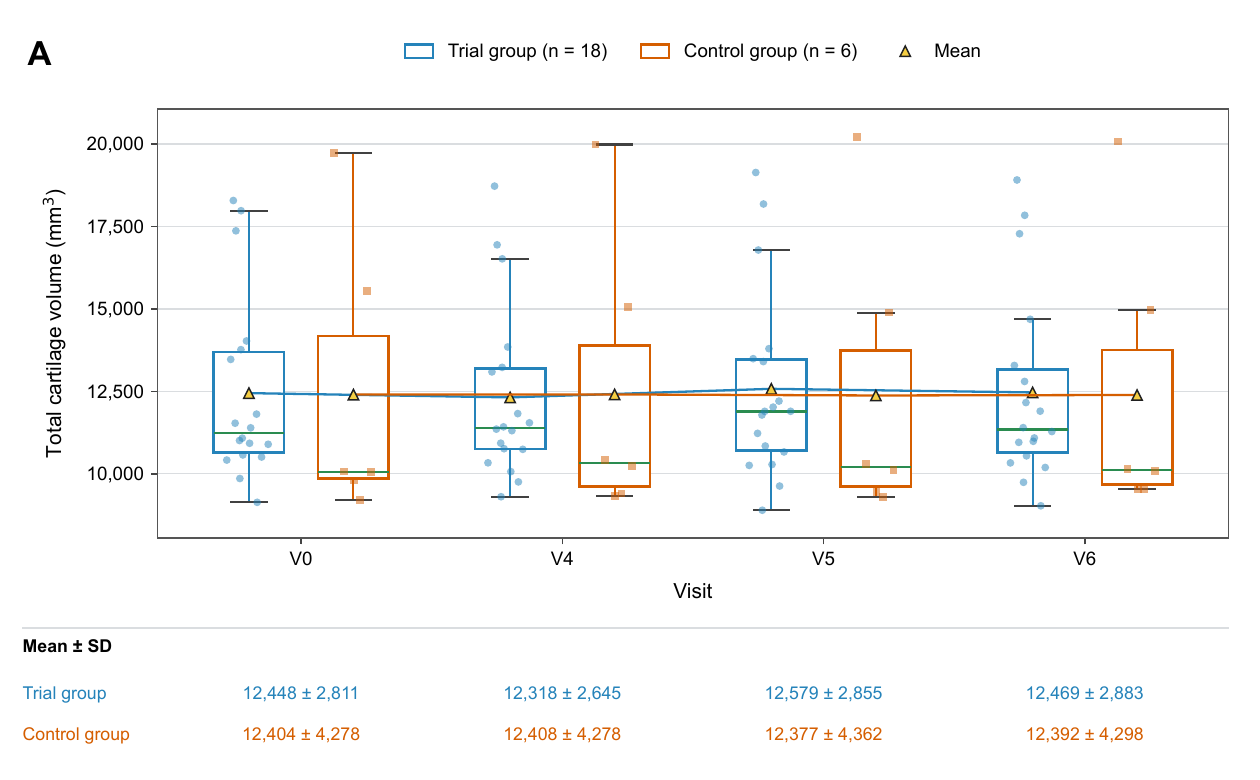}\par\vspace{4mm}
\includegraphics[width=0.94\linewidth]{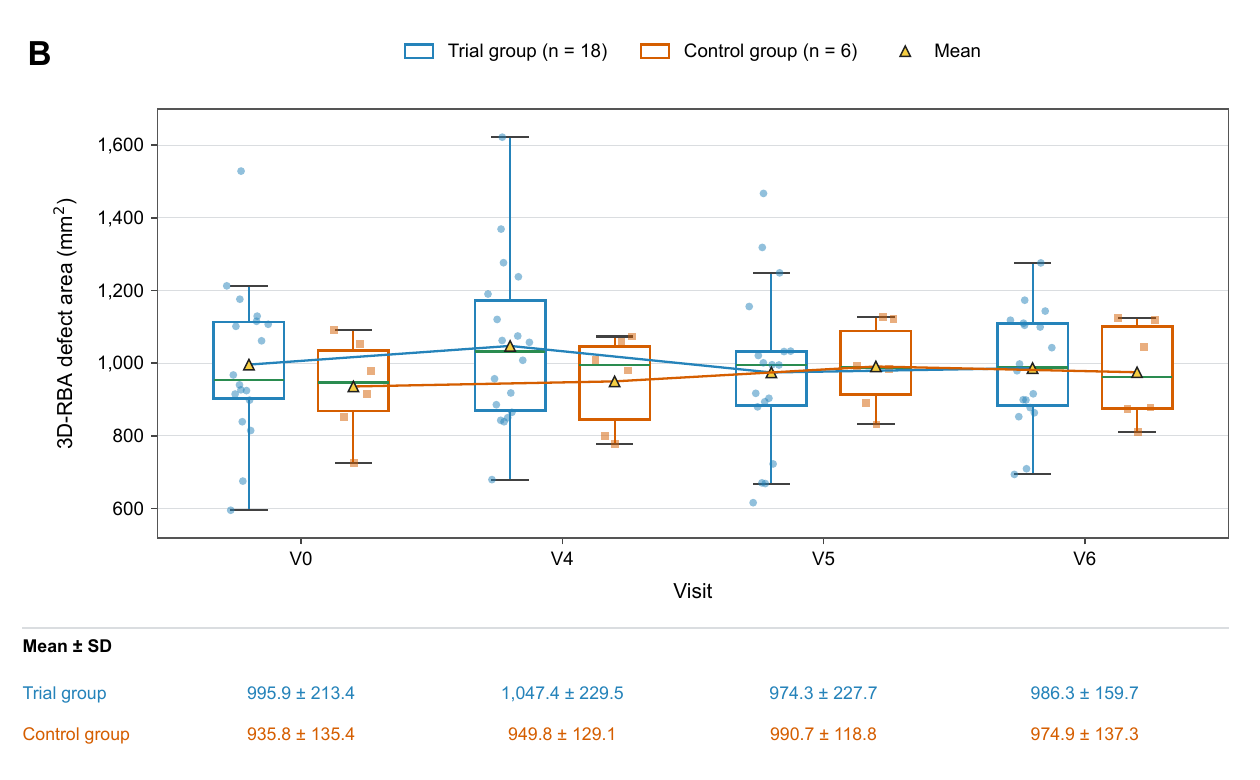}\par\medskip
\textbf{Figure 5. Phase II volume and defect area (A and B)}
\end{center}
\clearpage\begin{center}
\includegraphics[width=0.94\linewidth]{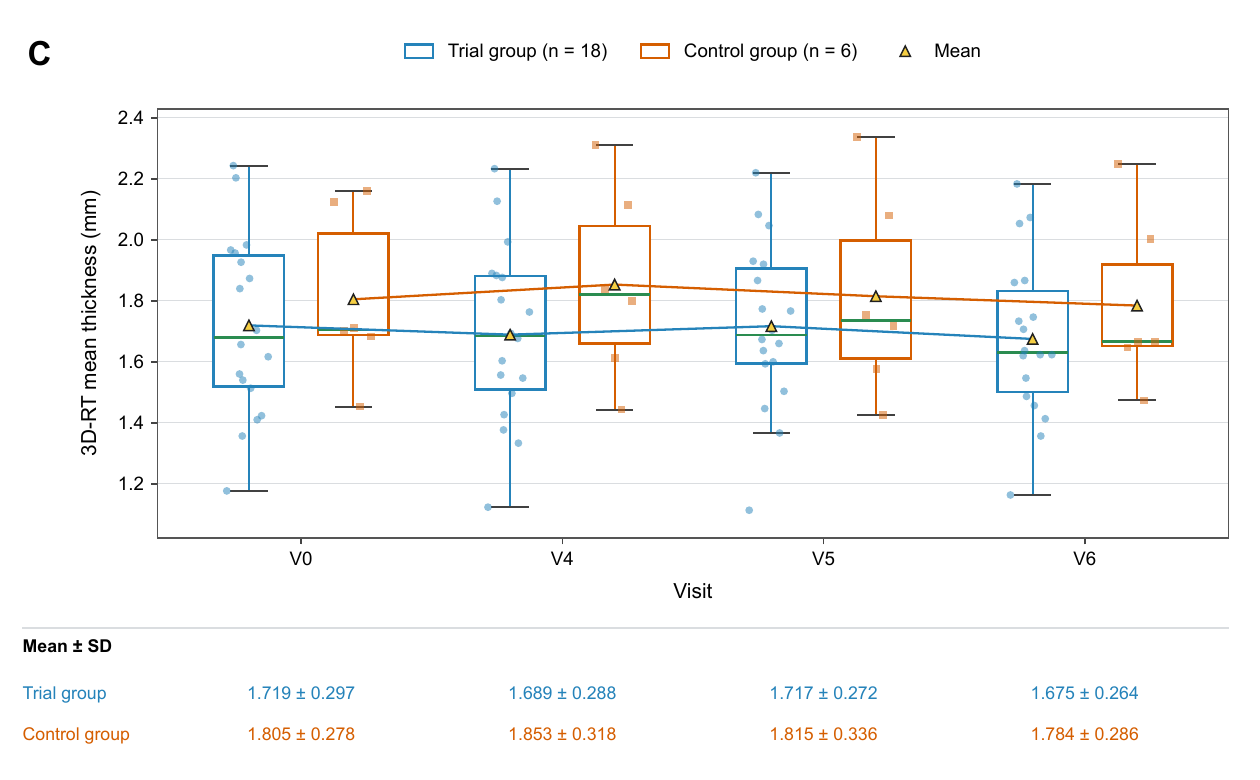}\par\medskip
\textbf{Figure 5 (continued). Phase II mean cartilage thickness (C)}
\end{center}
\clearpage
\setcounter{section}{0}
\setcounter{subsection}{0}
\setcounter{table}{0}
\setcounter{figure}{0}
\setcounter{equation}{0}
\renewcommand{\thesection}{S\arabic{section}}
\renewcommand{\thetable}{S\arabic{table}}
\renewcommand{\thefigure}{S\arabic{figure}}
\renewcommand{\theequation}{S\arabic{equation}}
\linespread{1.25}\selectfont

{\Large\bfseries Supporting information\par}
\medskip
{\large\bfseries Reliability assessment and multicenter clinical application of magnetic resonance methods for knee cartilage quantification\par}

This supplement provides eligibility criteria, acquisition parameters, model-development details, computational algorithms, statistical definitions, and extended numerical results. It accompanies the study design, measurement definitions, and principal findings in the main manuscript. Clinical metrics use femoral cartilage (FC), tibial cartilage (TC), and patellar cartilage (PC); Section S6 presents medial/lateral partitioning as a technical extension for regional analysis. Artificial intelligence (AI) and magnetic resonance imaging (MRI) are abbreviated as AI and MRI, respectively.

\section{Clinical imaging workflow and model development}
\subsection{Eligibility and additional protocol information}
Principal phase III inclusion criteria were knee osteoarthritis diagnosed according to the \textit{Guidelines for the diagnosis and treatment of osteoarthritis (2018 edition)}; age 40--80 years; body mass index $\leq30$ kg/m$^2$; knee pain lasting more than six months or recurrent/worsening symptoms after withdrawal of conventional treatment; Kellgren--Lawrence grade II--III in the study knee and $\leq$II in the contralateral knee; screening Western Ontario and McMaster Universities Osteoarthritis Index (WOMAC) total score of 50--150; study-knee visual analogue scale (VAS) score $\geq50$ mm on a 100-mm scale; voluntary written consent; and ability to complete the trial and follow-up. Principal exclusions included secondary knee osteoarthritis; other joint or systemic diseases affecting knee function; marked varus/valgus deformity ($>15^{\circ}$) or flexion contracture; previous stem cell treatment; intra-articular drug injection into either knee within six months before treatment; major surgery or arthroscopy of the study knee within 24 weeks before consent; planned relevant lower-limb surgery during the study; and MRI contraindications. The protocol also excluded recent analgesic, glucocorticoid, or other treatment potentially affecting efficacy assessment, severe comorbidities, pregnancy, and lactation.

Trial design, dosing, visit schedules, ethics approvals, and data authorization are reported in main Section 2.1; main Table 1 defines the analysis sets. The phase II study was a multicenter, randomized, double-blind, dose-finding trial of single-dose human umbilical cord mesenchymal stem cell injection (IxCell hUC-MSC-O). Retrospective annotation and quantification did not alter its intervention or follow-up. Phase II approval followed expedited review on December 3, 2020. Phase III underwent committee review on March 28, 2023, followed by reapproval on July 6, 2023; its protocol identifier was LC-MSCOA23001 and its registration was NCT06716281 (\url{https://clinicaltrials.gov/study/NCT06716281}).

\subsection{Acquisition requirements and quality control}
Under the common acquisition requirements described in main Section 2.2, the 14 centers used platform-adapted three-dimensional cartilage sequences, including the vendor sequence identifiers DE3D, Quick3D, VIBE, WATSc, MERGE, and FSPGR. Table~\ref{tab:acquisition} provides the charter's parameter ranges and requirements rather than observed scanner-frequency distributions. Quality checks included visit identity as well as laterality, coverage, motion artifacts, and sequence parameters.

\subsection{Model development and deployment}
All AI presegmentation models used the three-dimensional full-resolution configuration (\texttt{3d\_fullres}) of the self-configuring neural-network (nnU-Net) framework~\cite{s1}, with fully supervised learning from three-dimensional MRI and corresponding masks. nnU-Net performed experiment planning, resampling, and Z-score intensity normalization separately for each development dataset according to image resolution and dimensions. Models used a three-dimensional PlainConvUNet encoder--decoder, summed Dice and cross-entropy losses, deep supervision, and a batch size of two. He initialization was used. Optimization employed stochastic gradient descent with Nesterov momentum, an initial learning rate of 0.01, momentum of 0.99, weight decay of $3\times10^{-5}$, polynomial learning-rate decay, and a maximum of 1,000 epochs. Mirroring was disabled during training and inference for both V1.0 models; other default nnU-Net spatial and intensity augmentations were retained. V2.0 used standard nnU-Net augmentation and inference settings. Development data were organized into five folds; fold 0 was fixed for training and model selection, and weights were frozen before trial-image presegmentation.

The medical big-data training facility operated by the Shanghai Hospital Development Center (SHDC) aggregates multicenter clinical data through the Shanghai hospital information-sharing program. It provides data cleaning, de-identification, automatic or semiautomatic annotation, manual review, and model training and validation environments, with annotation quality-control mechanisms including single-reader review and two-reader annotation with adjudication~\cite{s2,s3}. Trial-image AI presegmentation and all subsequent two-reader correction and adjudication were performed within this facility.

V1.0 was deployed during the initial project stage. As neither OAI-ZIB nor SKI10 includes patellar cartilage labels, V1.0 comprised independent tibiofemoral and patellar cartilage models. The tibiofemoral development dataset included 726 examinations: 507 OAI-ZIB three-dimensional dual-echo steady-state (DESS) MRI examinations~\cite{s4}, 100 multivendor sagittal three-dimensional SKI10 examinations acquired at 1.0--3.0 T~\cite{s5}, and 119 knee MRI examinations from the SHDC facility. Fold 0 comprised 580 training and 146 within-fold validation examinations. The patellar model used the same 119 SHDC-facility examinations, with 95 for training and 24 for within-fold validation. The facility data included FC, TC, and PC labels, but these had been automatically generated by a third-party vendor model integrated into the facility and had not undergone the study's manual reference-standard workflow. They therefore served as non-reference-standard automatic annotations for development. The two models' outputs were combined at inference to form the V1.0 initial three-structure mask.

As manually annotated three-structure reference data accumulated, presegmentation was updated to V2.0. Its development-stage dataset comprised 428 knee MRI examinations from the SHDC facility with FC, TC, and PC reference labels obtained through standardized manual annotation and quality control. Of these, 400 entered the five-fold development pool; fold 0 contained 320 training and 80 within-fold validation examinations. A further 28 examinations formed a separate development-stage validation set, were not used to update model weights, and did not contribute validation results to this article's technical-performance analyses. V2.0 used one multiclass model to segment all three structures, aiming to improve training-label reliability, unify outputs, and reduce subsequent slice-by-slice manual correction.

Each version was frozen after training and internal validation and deployed according to project chronology: V1.0 for 700 and V2.0 for 489 phase III examinations, totaling 1,189. This complete deployment dataset supported agreement evaluation by comparing initial AI masks with adjudicated reference masks examination by examination. Internal-validation results were excluded from the technical-performance analyses. Development and internal-validation data did not overlap with these 1,189 trial examinations at either participant or examination level.

\subsection{Blinded reading and adjudication}
The blinded-reading procedure and reference-standard definition are described in main Section 2.2. The annotation system separately assigned each MRI and its initial AI mask to two readers and retained records of assignment, modification, resubmission, and adjudication. If neither submitted annotation was acceptable, the adjudicator documented the reason before returning both for relabeling. The adjudicated mask was then fixed for quantification. Intraclass correlation coefficients (ICCs) monitored continuous-metric agreement, including volume, during the trial.

Agreement analyses distinguished three data products: unmodified AI masks, each reader's independently corrected masks, and adjudicated reference masks. The primary quantitative agreement analysis compared the two readers, as specified in main Section 2.4; matches to adjudicated masks described the adjudication process rather than an independent-reader comparison.

All clinical metrics used consistent postprocessing methods and parameters. Reference-mask updates required by quality control were recalculated using the same workflow and checked before database lock. Geometric experiments and alternative-method analyses performed after database lock were supplementary methodological evaluations and did not alter the clinical quantification database. AI models generated initial segmentations only; all final metrics were calculated from manually adjudicated reference masks.

\section{Three-dimensional surfaces and quantification algorithms}
\subsection{Volume and thickness calculation}
Quantification used femoral cartilage (FC), tibial cartilage (TC), and patellar cartilage (PC) as three anatomical regions. Femoral cartilage excluded osteophytes and the intercondylar notch. Regional volume was $V=N\Delta x\Delta y\Delta z$, where $N$ is the target cartilage voxel count and $\Delta x$, $\Delta y$, and $\Delta z$ are physical voxel spacings. Total cartilage volume was the sum of the three regional volumes.

Marching Cubes extracted a triangular mesh at the binary mask's 0.5 isosurface~\cite{s6}. Meshes underwent 50 smoothing iterations with a relaxation factor of 0.1 and consistent normal orientation. The dot product of each vertex normal with the vector from that vertex to the corresponding bone centroid distinguished the cartilage--bone interface (inner surface) from the cartilage--joint-space interface (outer surface). Triangle area was $A=0.5\lVert(p_2-p_1)\times(p_3-p_1)\rVert$; deduplicated triangle areas were summed. All calculations used image physical coordinates.

Main Section 2.3 defines three-dimensional ray tracing (3D-RT), nearest neighbor (3D-NN), normal matching (3D-MN), and signed distance transform (3D-SDT). The implementations below specify sampling, missing-value handling, and aggregation for these thickness approaches~\cite{s7}, including Maurer et al.'s exact Euclidean distance transform~\cite{s8}. The same settings were used across visits. Clinical 3D-RT was also used for regional quantitative analyses submitted to the Center for Drug Evaluation (CDE), National Medical Products Administration.

\begin{algorithmblock}{3D-RT: ray-tracing thickness}
\textbf{Input:} Inner-surface vertices $P_{\mathrm{in}}$ and unit normals $n$; outer surface $S_{\mathrm{out}}$. Coordinates are in mm.
\algline{0}{$D_{\mathrm{RT}}\leftarrow$ empty sequence of thickness records.}
\algline{0}{\textbf{for} $p\in P_{\mathrm{in}}$:}
\algline{1}{$d\leftarrow$ local unit normal directed away from the bone centroid.}
\algline{1}{$Q\leftarrow$ intersections of $[p,p+100d]$ with $S_{\mathrm{out}}$.}
\algline{1}{\textbf{if} $Q=\varnothing$: append one zero to $D_{\mathrm{RT}}$.}
\algline{1}{\textbf{else}: $t\leftarrow\min_{q\in Q}\lVert q-p\rVert$; append $t$ to $D_{\mathrm{RT}}$ exactly $|Q|$ times.}
\algline{0}{$D_{\mathrm{valid}}\leftarrow$ finite records in $D_{\mathrm{RT}}$ satisfying $0\leq t\leq10$.}
\algline{0}{$T_{\mathrm{RT}}\leftarrow\operatorname{mean}(D_{\mathrm{valid}})$; if $D_{\mathrm{valid}}$ is empty but $D_{\mathrm{RT}}$ is not, use $\operatorname{median}(D_{\mathrm{RT}})$.}
\par\smallskip\textbf{Output:} Regional $T_{\mathrm{RT}}$, rounded to two decimal places.
\end{algorithmblock}
Zero values for absent intersections contribute to the regional mean. When a ray returns multiple intersections, the nearest distance is retained with multiplicity equal to the number of returned intersections, preserving the aggregation used for the study's actual ray-tracing results.

\subhead{Surface preparation for 3D-NN and 3D-MN}
Triangular meshes were extracted from the same cartilage masks and corresponding bone labels. Comparator cartilage meshes underwent 10 iterations of windowed-sinc smoothing with a passband of 0.12; bone meshes were not smoothed. Coordinates were transformed to image physical space. For each cartilage-surface point $p$, the nearest bone-surface point $b(p)$ and bone-to-cartilage unit vector $u(p)$ were determined. Inner and outer surfaces were classified using the relationship between the mesh normal and $u(p)$, distance to bone, and boundary-neighborhood assignments. Clearly oriented point sets were preferred; if either set contained fewer than 300 points, both reverted to the full classified sets. This yielded $P_{\mathrm{in}}$, $P_{\mathrm{out}}$, and $u(p)$ for outer-surface points.

The search limit was
\[
L=\min\{8,\max[2,1.2Q_{0.995}(d_b)]\}\ \mathrm{mm},
\]
where $d_b$ is cartilage-to-bone surface distance and $Q$ denotes an empirical quantile. The 3D-MN inward direction was $-u(p)$, not a direct substitution of the triangular-mesh normal.

\begin{algorithmblock}{3D-NN: nearest-neighbor thickness}
\textbf{Input:} Outer-surface points $P_{\mathrm{out}}$ and inner-surface points $P_{\mathrm{in}}$ in physical space.
\algline{0}{\textbf{for} $p\in P_{\mathrm{out}}$: $t_{\mathrm{NN}}(p)\leftarrow\min_{q\in P_{\mathrm{in}}}\lVert q-p\rVert$.}
\algline{0}{$D_{\mathrm{NN}}\leftarrow\{t_{\mathrm{NN}}(p):t_{\mathrm{NN}}(p)\text{ is finite and }>10^{-6}\ \mathrm{mm}\}$.}
\algline{0}{$n_{\mathrm{NN}}\leftarrow|D_{\mathrm{NN}}|$; $T_{\mathrm{NN}}\leftarrow\operatorname{TrimmedMean}(D_{\mathrm{NN}})$.}
\par\smallskip\textbf{Output:} Regional thickness $T_{\mathrm{NN}}$ and valid surface-point count $n_{\mathrm{NN}}$.
\end{algorithmblock}

\begin{algorithmblock}{3D-MN: normal-matching thickness}
\textbf{Input:} $P_{\mathrm{in}}$, $P_{\mathrm{out}}$, outward bone-reference unit vectors $u(p)$, and search limit $L$; $K=\min(128,|P_{\mathrm{in}}|)$.
\algline{0}{\textbf{for} $p\in P_{\mathrm{out}}$:}
\algline{1}{$d\leftarrow-u(p)$; $C(p)\leftarrow$ the $K$ nearest points in $P_{\mathrm{in}}$ to $p$.}
\algline{1}{For each $q\in C(p)$: $s(q)\leftarrow(q-p)\cdot d$;}
\algline{2}{$e(q)\leftarrow\lVert(q-p)-s(q)d\rVert$.}
\algline{1}{$C_{\mathrm{valid}}\leftarrow\{q\in C(p):0<s(q)\leq L\}$.}
\algline{1}{\textbf{if} $C_{\mathrm{valid}}\ne\varnothing$: choose $q^*$ minimizing $e(q)$; $t_{\mathrm{MN}}(p)\leftarrow s(q^*)$.}
\algline{1}{\textbf{else}: $t_{\mathrm{MN}}(p)\leftarrow0$.}
\algline{0}{$D_{\mathrm{MN}}\leftarrow\{t_{\mathrm{MN}}(p):t_{\mathrm{MN}}(p)\text{ is finite and }>10^{-6}\ \mathrm{mm}\}$.}
\algline{0}{$n_{\mathrm{MN}}\leftarrow|D_{\mathrm{MN}}|$; $T_{\mathrm{MN}}\leftarrow\operatorname{TrimmedMean}(D_{\mathrm{MN}})$.}
\par\smallskip\textbf{Output:} Regional thickness $T_{\mathrm{MN}}$ and valid surface-point count $n_{\mathrm{MN}}$.
\end{algorithmblock}

\begin{algorithmblock}{3D-SDT: signed-distance-transform thickness}
\textbf{Input:} Binary regional cartilage mask $M$ and physical voxel spacing $\Delta$.
\algline{0}{$D\leftarrow\operatorname{SignedMaurerDistanceMap}(M)$, with positive internal distances, nonsquared distances, and physical voxel spacing.}
\algline{0}{\textbf{for} $x$ with $M(x)=1$: $t_{\mathrm{SDT}}(x)\leftarrow2D(x)$.}
\algline{0}{$D_{\mathrm{SDT}}\leftarrow\{t_{\mathrm{SDT}}(x):t_{\mathrm{SDT}}(x)\text{ is finite and }>10^{-6}\ \mathrm{mm}\}$.}
\algline{0}{$T_{\mathrm{SDT}}\leftarrow\operatorname{TrimmedMean}(D_{\mathrm{SDT}})$;}
\algline{0}{$V\leftarrow$ cartilage voxel count $\times\Delta x\Delta y\Delta z$.}
\par\smallskip\textbf{Output:} Regional thickness $T_{\mathrm{SDT}}$ and cartilage volume $V$.
\end{algorithmblock}
3D-SDT samples internal cartilage voxels; it does not extract a medial axis or maximal inscribed spheres. Zero distances at the boundary do not contribute to its regional mean.

\subhead{Valid-value processing and three-region aggregation}
For $\operatorname{TrimmedMean}(D)$, return missing if $D$ is empty, and return $\operatorname{mean}(D)$ if $|D|<50$. Otherwise, set $a=Q_{0.01}(D)$ and $b=Q_{0.99}(D)$ and average records satisfying $a\leq t\leq b$. If trimming leaves no records, return $\operatorname{mean}(D)$.
\begin{align*}
T_{\mathrm{RT,overall}}&=\frac{T_{\mathrm{RT,FC}}+T_{\mathrm{RT,TC}}+T_{\mathrm{RT,PC}}}{3},\\
T_{\mathrm{method,overall}}&=\frac{\sum_r n_{\mathrm{method},r}T_{\mathrm{method},r}}{\sum_r n_{\mathrm{method},r}},\quad \mathrm{method}\in\{\mathrm{3D\mbox{-}NN},\mathrm{3D\mbox{-}MN}\},\\
T_{\mathrm{SDT,overall}}&=\frac{\sum_r V_rT_{\mathrm{SDT},r}}{\sum_r V_r},\quad r\in\{\mathrm{FC},\mathrm{TC},\mathrm{PC}\}.
\end{align*}
3D-NN and 3D-MN weights are the positive, finite surface-point counts \emph{before} trimming; 3D-SDT weights are volumes of all cartilage voxels in each region. Trimming applies only to 3D-NN, 3D-MN, and 3D-SDT regional means, not clinical 3D-RT. Each method retains its measurement definition and sampling weights for longitudinal comparison in the same participants; values are not rescaled to enforce agreement beyond using common units.

\subsection{Ray-based defect-area implementation}
Three-dimensional ray-based area (3D-RBA) implements the 1.5-mm operational threshold defined in main Section 2.3. Its rationale draws on International Cartilage Repair Society (ICRS) lesion-depth grading~\cite{s9} and healthy tibiofemoral thickness measurements~\cite{s10,s11,s12}, not a patient-specific conversion to an ICRS grade. Previous MRI coverage studies provide context for reproducible surface-area measurements~\cite{s13,s14,s15}. The algorithm below specifies the two contributing triangle sets, physical-coordinate area calculation, and deduplication.

\begin{algorithmblock}{3D-RBA: ray-based defect-area estimation}
\textbf{Input:} Adjudicated cartilage mask $M$, image-to-physical coordinate transform $T$, and thickness threshold $\tau=1.5$ mm.
\algline{0}{$(S_{\mathrm{in}},S_{\mathrm{out}})\leftarrow$ reconstruct a triangular mesh from $M$ and identify inner and outer cartilage surfaces.}
\algline{0}{$A_{\mathrm{thin}}\leftarrow\varnothing$; $A_{\mathrm{zero}}\leftarrow\varnothing$.}
\algline{0}{\textbf{for} $p\in\operatorname{vertices}(S_{\mathrm{in}})$:}
\algline{1}{Cast a ray along the local normal at $p$; find its first valid intersection $q$ with $S_{\mathrm{out}}$.}
\algline{1}{\textbf{if} $q$ exists:}
\algline{2}{$t(p)\leftarrow\lVert T(q)-T(p)\rVert$.}
\algline{2}{\textbf{if} $0<t(p)<\tau$: add the outer-surface triangles adjacent to $q$ to $A_{\mathrm{thin}}$.}
\algline{1}{\textbf{else}:}
\algline{2}{$t(p)\leftarrow0$.}
\algline{2}{Add the inner-surface triangles specified by the boundary-compensation rules to $A_{\mathrm{zero}}$.}
\algline{0}{$A\leftarrow$ deduplicate $A_{\mathrm{thin}}\cup A_{\mathrm{zero}}$.}
\par\smallskip\textbf{Output:} $\mathrm{3D\mbox{-}RBA}=\sum_{f\in A}\operatorname{Area}_{T}(f)$.
\end{algorithmblock}
$A_{\mathrm{thin}}$ contains regions with residual cartilage below the threshold, representing partial-thickness defects and diffuse thinning. $A_{\mathrm{zero}}$ contains regions with no detected valid cartilage thickness, treated as zero to estimate candidate full-thickness defects.

Physiologically thin regions and natural cartilage margins remain subject to the same threshold and boundary-compensation rules. Previous histological and arthroscopic comparisons provide clinical context for MRI lesion quantification~\cite{s16,s17,s18}; they do not supply a calibration of this implementation. The present method's evaluation and interpretation are reported in the main manuscript.

\subsection{Alternative defect-area estimation}
Three-dimensional patch-mapped area (3D-PMA) assessed sensitivity of longitudinal results to defect-area definition. The implementation drew on MRI cartilage-coverage, bone-reference surface-mapping, and threshold-based surface-area studies~\cite{s19,s13,s14,s20,s21} to estimate denuded bone surface area (dAB), a measure used in osteoarthritis imaging research~\cite{s22}. Its reference surface and classification rules differ from those of 3D-RBA. The adaptive threshold uses the median absolute deviation (MAD) as one component of its robust distance rule. It serves as an alternative-definition sensitivity analysis, not a calibration standard for absolute 3D-RBA values.

\begin{algorithmblock}{3D-PMA: patch-mapped defect-area estimation}
\textbf{Input:} Bone surface $B$ and inner cartilage-surface points $C$ corresponding to FC, TC, and PC, in physical coordinates.
\algline{0}{\textbf{for each} anatomical region:}
\algline{1}{$d(b)\leftarrow$ nearest distance from each bone-surface vertex $b$ to $C$.}
\algline{1}{$P\leftarrow$ bone-surface patch adjacent to cartilage, identified using distance, normal relationships, and connectivity.}
\algline{1}{$D\leftarrow$ valid distances meeting normal criteria and within the search limit.}
\algline{1}{$\tau_c\leftarrow\max\{\operatorname{Otsu}(D),\operatorname{median}(D_{\mathrm{low20\%}})+3\times1.4826\times\mathrm{MAD}\}$~\cite{s23}.}
\algline{1}{Constrain $\tau_c$ to prespecified bounds; use a voxel-spacing-dependent fallback when samples are insufficient.}
\algline{1}{\textbf{for each} triangle $f$ with all three vertices in $P$:}
\algline{2}{Linearly interpolate vertex distances and calculate physical area $A_f$ where $d>\tau_c$.}
\par\smallskip\textbf{Output:} $\mathrm{3D\mbox{-}PMA}=\sum A_f$ over all three regions (\mmtwo).
\end{algorithmblock}
MAD denotes median absolute deviation; $\tau_c$ is the bone--cartilage contact threshold, and $D_{\mathrm{low20\%}}$ denotes the lowest 20\% of distance samples.

\section{Geometric references and statistical methods}
Forty participants were selected by simple random sampling without replacement from the 416-participant phase III imaging cohort (random seed, 20260929), before geometric measurements. Baseline FC, TC, and PC masks yielded 120 participant--region experiments, with no replacements or exclusions. Each mask was used to create a 3-mm baseline and a continuous, approximately 100-\mmtwo\ region with 1-mm residual cartilage. The actual reference area was not assumed to equal the nominal design area: a Euclidean distance field from the bone mask independently identified subthreshold regions within the predefined design footprint on the reconstructed surface. Triangle areas in these regions were included when the sampled distance was $<1.5$ mm. Spatial agreement was evaluated in the design footprint and its 2-mm neighborhood on the same reconstructed surface.

For each post-thinning cartilage surface, the nearest valid intersection along a surface normal defined local ray thickness. Locations with $0<t<1.5$ mm were classified as thinned, and their associated triangles were deduplicated before summing physical surface areas. Locations without a valid intersection were not included in this controlled partial-thickness experiment. The geometric estimate was the total post-thinning selected area, compared with the independent distance-field reference.

Evaluation included bias, mean absolute error (MAE), root mean square error (RMSE), mean absolute percentage error (MAPE), Pearson and Spearman correlations, Lin's concordance correlation coefficient (CCC)~\cite{s25}, and area-weighted precision, recall, Dice, and intersection over union (IoU). The latter compared selected and independently labeled subthreshold triangles in the 2-mm design-local domain. The 95\% confidence intervals (CIs) were percentile intervals from 2,000 bootstrap resamples of participants, retaining all three regions for each sampled participant. Figure S5 shows experiment-level spatial distributions, and Figure S6 shows absolute percentage errors for all 120 participant--region experiments.

Continuous variables are expressed as mean $\pm$ standard deviation or median (interquartile range, range), and categorical variables as counts and percentages. With adjudicated masks as reference, Dice, IoU, 95th-percentile Hausdorff distance (HD95), and average surface distance (ASD) were calculated for FC, TC, PC, and their union. For predicted mask $P$ and reference mask $G$,
\[
\mathrm{Dice}=\frac{2|P\cap G|}{|P|+|G|},\qquad
\mathrm{IoU}=\frac{|P\cap G|}{|P\cup G|}.
\]
Dice and IoU quantify voxel overlap. For nonempty masks, let $S_P=\{p_i\}_{i=1}^{n_P}$ and $S_G=\{g_j\}_{j=1}^{n_G}$ denote the predicted and reference surface voxel centers, respectively, extracted using six-connected neighborhoods and transformed into Neuroimaging Informatics Technology Initiative (NIfTI) physical coordinates. The directional nearest-surface distances are
\[
d_i=\min_{1\leq j\leq n_G}\lVert p_i-g_j\rVert_2,
\qquad
e_j=\min_{1\leq i\leq n_P}\lVert g_j-p_i\rVert_2.
\]
The surface-distance metrics were calculated as
\begin{align*}
\mathrm{HD95}&=Q_{0.95}\bigl((d_1,\ldots,d_{n_P},e_1,\ldots,e_{n_G})\bigr),\\
\mathrm{ASD}&=\frac{1}{2}\left(
\frac{1}{n_P}\sum_{i=1}^{n_P}d_i+
\frac{1}{n_G}\sum_{j=1}^{n_G}e_j\right).
\end{align*}
Here, $n_P$ and $n_G$ are the respective surface voxel counts, $\lVert\cdot\rVert_2$ is the Euclidean norm, and $Q_{0.95}$ denotes the empirical 95th percentile. HD95 pools all distance samples from both directions, retaining repeated and zero distances; ASD gives equal weight to the two directional means. Both metrics are in mm, with lower values indicating closer boundaries. The same definitions were applied to interreader mask comparisons.
For two empty masks, Dice and IoU were set to 1 and HD95 and ASD to 0; when exactly one mask was empty, HD95 and ASD were treated as missing.

The ICC model, resampling units, and iteration counts are specified in main Section 2.4. Participant-bootstrap CIs used the 2.5th and 97.5th percentiles. For repeated-measures Bland--Altman analysis~\cite{s30}, let $d_{iv}=A_{iv}-B_{iv}$ denote the difference between Reader A and Reader B for participant $i$ at examination $v$. A random-intercept variance decomposition gave participant and residual components, $\sigma_u^2$ and $\sigma_e^2$. The total difference standard deviation was $\sigma_d=\sqrt{\sigma_u^2+\sigma_e^2}$, and the 95\% LoA were $\bar d\pm1.96\sigma_d$. The interreader SEM was $\sigma_d/\sqrt{2}$ and MDC95 was $1.96\sqrt{2}\times\mathrm{SEM}$~\cite{s31}. These MDC95 estimates quantify measurement error attributable to reader-dependent mask correction within this workflow. CIs used 2,000 participant-level bootstrap iterations with all examinations retained for each sampled participant. Proportional bias was assessed by regression of $d_{iv}$ on the paired mean, using participant-clustered standard errors. Phase III correlations used one final-minus-baseline change per participant for each metric; all evaluated pairs are listed in main Table 3. Main Figure 4A,B presents visit distributions, and Figure 4C combines complete paired trajectories for total volume, 3D-RT, and 3D-RBA in three plots for the same 65 participants. Percentage changes were calculated from visit means, not by averaging individual percentage changes.

For phase II metric $x$, participant-centered values were
\[
x^{c}_{iv}=x_{iv}-\frac{1}{4}\sum_{u\in\{\mathrm{V0,V4,V5,V6}\}}x_{iu},
\]
where $i$ denotes the participant and $v$ the visit. Pearson correlations were computed between centered metric pairs separately within treatment and control groups. Entire participant clusters, each retaining all four visits, were resampled for CIs. This removes participant-specific means but does not adjust for visit effects. Tables~\ref{tab:phaseii} and~\ref{tab:within} report unadjusted visit summaries and participant-centered correlations, respectively.

Previous validation studies using synthetic models and three-dimensional cartilage morphometry provide methodological context for controlled geometric evaluation~\cite{s26,s27}. The present geometric errors characterize performance under predefined contiguous thinning, whereas clinical longitudinal analyses examine change relationships across metrics and definitions.

\section{Supplementary segmentation agreement and geometric results}
Main Section 3.1 reports AI presegmentation agreement with adjudicated masks and interreader agreement. Across the 1,189 deployment examinations, median overall Dice was 0.971; 942 examinations (79.2\%) had Dice $\geq0.95$, 49 had Dice $<0.90$, and 39 had Dice of 1.000. Table~\ref{tab:segmentation} provides the complete structure-specific overlap and surface-distance summaries underlying main Figure 1.

V1.0 was deployed in 700 examinations, with overall Dice, IoU, HD95, and ASD of $0.967\pm0.017$, $0.937\pm0.031$, $0.637\pm0.617$ mm, and $0.104\pm0.070$ mm, respectively. For the 489 V2.0 examinations, corresponding values were $0.960\pm0.040$, $0.926\pm0.072$, $0.384\pm0.534$ mm, and $0.089\pm0.147$ mm. These chronologically determined subsets were not used to infer version superiority. Principal-visit distributions overlapped; one nonprincipal examination was excluded from visit-stratified figures.

Figures S1--S4 show overall frequencies, principal-visit distributions, and version-specific agreement. Main Figure 2 presents representative three-dimensional contours of the adjudicated cartilage structures. Figures S5 and S6 provide experiment-level spatial results and an error heatmap, complementing the aggregate geometric assessment in main Figure 3.

Repeated-measures Bland--Altman analysis of the two independent readers is summarized in Table~\ref{tab:absolute-agreement} and Figure S7. Overall biases were 367.7 \mmthree, 0.030 mm, and 6.9 \mmtwo\ for volume, 3D-RT, and 3D-RBA, respectively. Their corresponding MDC95 values were 1,596.9 \mmthree, 0.227 mm, and 147.4 \mmtwo. Proportional-bias slopes were statistically detectable for all three overall measures ($P<0.001$), but explained only 3.7\%, 4.6\%, and 1.9\% of difference variance, respectively. Region-specific absolute-agreement estimates are provided in Table~\ref{tab:absolute-agreement}.

Table~\ref{tab:geometry} extends the principal geometric findings in main Section 3.2 with the complete error and overlap metrics. RMSE was 5.363 \mmtwo; Pearson and Spearman correlations were 0.907 and 0.921. Spatial precision was 0.997, recall 0.929, and IoU 0.926.

\section{Extended clinical longitudinal results}
\subsection{Three-visit phase III metric distributions}
Main Table 1 defines the analysis sets, and main Table 3 reports paired-change correlations in the 374 complete cases and the 65-participant increasing-volume subset. The visit-level values below provide numerical summaries for main Figure 4, using the same 65 participants at baseline (V0), week 24 (V6), and week 48 (V8).

In main Figure 4A, total cartilage volume at V0, V6, and V8 was $14,160.848\pm2,911.433$, $14,484.751\pm2,956.614$, and $15,192.085\pm3,184.911$ \mmthree, respectively, a 7.28\% V8 increase from V0. Corresponding 3D-RBA values were $987.922\pm187.726$, $991.291\pm214.880$, and $938.334\pm223.706$ \mmtwo. The V6 mean was close to baseline, and V8 was 5.02\% lower. 3D-PMA was $1,200.455\pm313.698$, $1,188.980\pm347.414$, and $1,181.399\pm327.372$ \mmtwo, with consecutive decreases and a 1.59\% V8 reduction. Under progressively increasing total volume, final mean defect area was below baseline with both definitions.

In main Figure 4B, 3D-RT mean thickness at V0, V6, and V8 was $2.189\pm0.252$, $2.195\pm0.262$, and $2.264\pm0.271$ mm. Corresponding values were $1.376\pm0.148$, $1.391\pm0.126$, and $1.464\pm0.161$ mm for 3D-NN; $1.351\pm0.156$, $1.368\pm0.127$, and $1.461\pm0.173$ mm for 3D-MN; and $0.854\pm0.118$, $0.860\pm0.121$, and $0.884\pm0.124$ mm for 3D-SDT. All four mean thicknesses increased across visits; V8 increases were 3.39\%, 6.43\%, 8.11\%, and 3.53\%, respectively. Main Section 3.3 reports measurement completeness, individual trajectory summaries, and correlations with alternative estimates.

\subsection{Four-visit phase II results}
Table~\ref{tab:phaseii} reports the complete numerical results for main Figure 5: 18 treated participants and six controls, each observed at screening baseline (V0) and post-dose days 29 (V4), 85 (V5), and 169 (V6). Treatment-group mean 3D-RBA rose at V4 and fell at V5; its final value was below baseline, whereas the control-group final mean exceeded baseline. The full visit summaries retain the nonmonotonic patterns across metrics.

Table~\ref{tab:within} provides all six within-participant correlations and their CIs, complementing main Section 3.4. Volume--thickness and volume--defect-area relationships had the same respective directions in both groups; defect-area--thickness CIs crossed zero in both groups. These results use participant-centered observations rather than correlations between group means.

\section{Technical extension and discussion of medial/lateral stratification}
Regional heterogeneity of knee osteoarthritis supports finer spatial characterization through medial/lateral stratification. Within this clinical imaging workflow, adjudicated FC, TC, and PC masks were further partitioned into medial and lateral femoral, medial and lateral tibial, and patellar regions for stratified statistics and regional quantitative results submitted to CDE.

Training labels were extended from the public OAI-ZIB knee MRI dataset~\cite{s4} using the CLAIR-Knee-103R cartilage atlas~\cite{s28}. OAI-ZIB provides whole femoral cartilage and already separated medial and lateral tibial cartilage, retained as medial tibial cartilage (MTC) and lateral tibial cartilage (LTC). For unpartitioned femoral cartilage, the CLAIR-Knee-103R template and its 20-region atlas were registered to individual space. Corresponding atlas subregions were merged to derive medial femoral cartilage (MFC) and lateral femoral cartilage (LFC) reference regions. MFC--LFC interface points were transformed into physical coordinates, and orthogonal least-squares plane fitting based on singular value decomposition (SVD)~\cite{s29} defined a plane for splitting the original manual femoral cartilage mask. Atlas information determined orientation while preserving manual contours, yielding five training labels: background, MFC, LFC, MTC, and LTC.

The partitioning model used nnU-Net v2's three-dimensional full-resolution configuration~\cite{s1}. Images were resampled to 1.50-mm through-plane and $0.3125\times0.3125$-mm in-plane spacing and Z-score normalized. Patch size was $32\times256\times224$ voxels and batch size two. A six-stage three-dimensional PlainConvUNet used summed Dice and cross-entropy losses with deep supervision. Data were split into five folds; fold 0 was fixed for training and inference. Stochastic gradient descent used Nesterov momentum, initial learning rate 0.01, momentum 0.99, weight decay $3\times10^{-5}$, polynomial decay, and 1,000 epochs. Training and test-time mirroring were disabled to preserve medial/lateral label semantics; other spatial and intensity augmentations followed nnU-Net defaults.

At application, predicted MFC, LFC, MTC, and LTC determined anatomical boundaries. Adjacent MFC--LFC interface points were extracted and transformed into physical space. Their centroid defined the plane's reference point. SVD of the centered-coordinate matrix yielded the right singular vector associated with the smallest singular value as the normal, whose medial/lateral orientation was standardized using an MFC reference point. Signed physical distance to this plane partitioned adjudicated femoral and tibial cartilage masks; PC remained separate. Corresponding bone labels followed the same boundary rule to preserve regional bone--cartilage correspondence. Volume, 3D-RT, and 3D-RBA were then calculated regionally. Partitioning changed anatomical label assignments without altering cartilage contours, and physicians visually reviewed the results.

This strategy combines atlas-based anatomical localization with adjudicated-mask contours to define consistent regional analysis boundaries while retaining cartilage morphology. Visual review identifies conspicuous boundary or anatomical-assignment errors; quantitative boundary accuracy and repeatability require further evaluation against independent partitioning reference standards.

The principal methodological results in this article use the three adjudicated FC, TC, and PC regions. Five-region partitioning is a technical extension providing a reproducible implementation basis for future regional metric evaluation.

\clearpage
\begingroup\linespread{1.0}\selectfont\small

\par\endgroup

\clearpage
\begingroup\linespread{1.0}\selectfont\small
\refstepcounter{table}\label{tab:acquisition}
\textbf{Table \thetable. Phase III three-dimensional knee cartilage MRI acquisition requirements}
\par\smallskip
\begin{tabularx}{\textwidth}{@{}p{49mm}Y@{}}
\toprule
Item & Range/requirement \\
\midrule
Field strength & 1.5 or 3.0 T \\
Coil & Knee coil with at least four channels \\
Sequences & DE3D / Quick3D / VIBE / WATSc / MERGE / FSPGR, etc. \\
Matrix & $256\times256$ to $512\times512$ \\
Slice thickness & 1.5 mm \\
In-plane pixel size & 0.22--0.35 mm \\
Repetition time (TR) / echo time (TE) & 3--30 ms / 5--17 ms \\
Field of view (FOV) & 14--18 cm \\
Number of slices & 58--112 \\
Flip angle & $10^{\circ}$--$90^{\circ}$ \\
Number of excitations (NEX) & 1--3 \\
\bottomrule
\end{tabularx}
\par\smallskip\footnotesize
Field strength, coils, and imaging parameters represent project requirements or recommended ranges; sequence names summarize acquisition records from 14 centers. Each center followed its project-approved site protocol. Equipment, coil, and sequence parameters were kept consistent across visits whenever possible. TR, repetition time; TE, echo time; FOV, field of view; NEX, number of excitations.
\par\endgroup

\clearpage
\begingroup\linespread{1.0}\selectfont\small
\refstepcounter{table}\label{tab:segmentation}
\textbf{Table \thetable. Spatial agreement between initial AI and adjudicated reference masks}
\par\smallskip
\begin{tabularx}{\textwidth}{@{}Ycccc@{}}
\toprule
Target & Dice & IoU & HD95 (mm) & ASD (mm) \\
\midrule
Overall & $0.964\pm0.029$ & $0.932\pm0.052$ & $0.533\pm0.597$ & $0.098\pm0.109$ \\
Femoral cartilage & $0.972\pm0.027$ & $0.947\pm0.048$ & $0.319\pm0.549$ & $0.079\pm0.074$ \\
Tibial cartilage & $0.959\pm0.042$ & $0.924\pm0.072$ & $1.099\pm1.715$ & $0.152\pm0.209$ \\
Patellar cartilage & $0.935\pm0.068$ & $0.885\pm0.106$ & $0.755\pm2.276$ & $0.133\pm0.372$ \\
\bottomrule
\end{tabularx}
\par\smallskip\footnotesize
Values are mean $\pm$ standard deviation. Dice and IoU range from 0 to 1; higher values indicate greater overlap. HD95 and ASD are in mm; lower values indicate closer surfaces. Overall cartilage is the union of the femoral, tibial, and patellar masks.
\par\endgroup

\clearpage
\begingroup\linespread{1.0}\selectfont\small
\refstepcounter{table}\label{tab:geometry}
\textbf{Table \thetable. Three-region digital-phantom evaluation in 120 participant--region experiments}
\par\smallskip
\begin{tabularx}{\textwidth}{@{}Yp{71mm}@{}}
\toprule
Metric & Value \\
\midrule
Mean reference / measured post-thinning defect area & 86.793 / 83.675 \mmtwo \\
Bias & $-3.119$ \mmtwo \\
MAE (95\% CI) & 4.627 (4.180--5.113) \mmtwo \\
RMSE & 5.363 \mmtwo \\
MAPE (95\% CI) & 5.616\% (4.946\%--6.457\%) \\
Pearson / Spearman correlation & 0.907 / 0.921 \\
CCC & 0.863 \\
Precision / recall & 0.997 / 0.929 \\
Dice / IoU & 0.961 / 0.926 \\
\bottomrule
\end{tabularx}
\par\smallskip\footnotesize
MAE, mean absolute error; RMSE, root mean square error; MAPE, mean absolute percentage error; CCC, concordance correlation coefficient; IoU, intersection over union; CI, confidence interval. Error metrics compare post-thinning total area with independently labeled subthreshold area in the predefined design footprint. Spatial metrics are case-level means from the 2-mm design-local domain.
\par\endgroup

\clearpage
\begingroup\linespread{1.0}\selectfont\small
\refstepcounter{table}\label{tab:phaseii}
\textbf{Table \thetable. Cartilage metrics in the complete four-visit phase II cohort}
\par\smallskip
\begin{tabularx}{\textwidth}{@{}Yccc@{}}
\toprule
Metric & Visit & Treatment ($n=18$) & Control ($n=6$) \\
\midrule
Total cartilage volume (\mmthree) & V0 & $12,448\pm2,811$ & $12,404\pm4,278$ \\
 & V4 & $12,318\pm2,645$ & $12,408\pm4,278$ \\
 & V5 & $12,579\pm2,855$ & $12,377\pm4,362$ \\
 & V6 & $12,469\pm2,883$ & $12,392\pm4,298$ \\
\midrule
3D-RBA estimated cartilage defect area (\mmtwo) & V0 & $995.9\pm213.4$ & $935.8\pm135.4$ \\
 & V4 & $1,047.4\pm229.5$ & $949.8\pm129.1$ \\
 & V5 & $974.3\pm227.7$ & $990.7\pm118.8$ \\
 & V6 & $986.3\pm159.7$ & $974.9\pm137.3$ \\
\midrule
3D-RT mean cartilage thickness (mm) & V0 & $1.719\pm0.297$ & $1.805\pm0.278$ \\
 & V4 & $1.689\pm0.288$ & $1.853\pm0.318$ \\
 & V5 & $1.717\pm0.272$ & $1.815\pm0.336$ \\
 & V6 & $1.675\pm0.264$ & $1.784\pm0.286$ \\
\bottomrule
\end{tabularx}
\par\smallskip\footnotesize
The same 24 participants contributed observations at V0, V4, V5, and V6, totaling 96 examinations. V0 denotes screening baseline; V4, V5, and V6 denote post-dose days 29, 85, and 169. Volume and 3D-RBA are sums across FC, TC, and PC; 3D-RT is the unweighted arithmetic mean of regional means. Values are unadjusted means $\pm$ standard deviations.
\par\endgroup

\clearpage
\begingroup\linespread{1.0}\selectfont\small
\refstepcounter{table}\label{tab:within}
\textbf{Table \thetable. Within-participant correlations across four phase II visits}
\par\smallskip
\begin{tabularx}{\textwidth}{@{}Ycc@{}}
\toprule
Metric pair & Treatment $r$ (95\% CI) & Control $r$ (95\% CI) \\
\midrule
Volume and 3D-RT & 0.529 (0.322 to 0.707) & 0.762 (0.689 to 0.868) \\
Volume and 3D-RBA & $-0.429$ ($-0.612$ to $-0.253$) & $-0.334$ ($-0.537$ to $-0.131$) \\
3D-RBA and 3D-RT & $-0.214$ ($-0.478$ to 0.018) & $-0.355$ ($-0.701$ to 0.063) \\
\bottomrule
\end{tabularx}
\par\smallskip\footnotesize
Treatment: 18 participants; control: six participants. Pearson correlations were calculated after subtracting each participant's four-visit mean from each metric. CIs used 5,000 whole-participant bootstrap iterations. Visit effects were not adjusted, and these analyses were not used for between-group efficacy testing.
\par\endgroup

\clearpage
\begingroup\linespread{1.0}\selectfont\footnotesize
\refstepcounter{table}\label{tab:absolute-agreement}
\textbf{Table \thetable. Repeated-measures Bland--Altman agreement and interreader MDC95}
\par\smallskip
\begin{tabularx}{\textwidth}{@{}>{\raggedright\arraybackslash}p{29mm}>{\raggedright\arraybackslash}p{21mm}YYY@{}}
\toprule
Region & Metric & Bias (95\% CI) & 95\% LoA & MDC95 (95\% CI) \\
\midrule
Overall cartilage & Volume (\mmthree) & 367.7 (318.2 to 419.4) & $-1,229.1$ to 1,964.6 & 1,596.9 (1,478.9 to 1,715.9) \\
 & 3D-RT (mm) & 0.030 (0.022 to 0.038) & $-0.197$ to 0.257 & 0.227 (0.213 to 0.239) \\
 & 3D-RBA (\mmtwo) & 6.9 (2.7 to 11.4) & $-140.5$ to 154.3 & 147.4 (132.6 to 163.8) \\
\midrule
Femoral cartilage & Volume (\mmthree) & 111.7 (84.4 to 140.0) & $-816.7$ to 1,040.1 & 928.4 (834.9 to 1,035.9) \\
 & 3D-RT (mm) & 0.010 (0.006 to 0.013) & $-0.091$ to 0.111 & 0.101 (0.088 to 0.115) \\
 & 3D-RBA (\mmtwo) & $-1.3$ ($-4.6$ to 2.0) & $-115.3$ to 112.7 & 114.0 (103.0 to 126.5) \\
\midrule
Tibial cartilage & Volume (\mmthree) & 123.7 (111.4 to 137.0) & $-305.9$ to 553.4 & 429.6 (382.1 to 484.3) \\
 & 3D-RT (mm) & $-0.033$ ($-0.039$ to $-0.028$) & $-0.208$ to 0.141 & 0.174 (0.157 to 0.193) \\
 & 3D-RBA (\mmtwo) & 12.6 (10.9 to 14.4) & $-43.5$ to 68.7 & 56.1 (47.8 to 65.1) \\
\midrule
Patellar cartilage & Volume (\mmthree) & 132.3 (108.8 to 157.1) & $-581.6$ to 846.3 & 714.0 (674.7 to 753.8) \\
 & 3D-RT (mm) & 0.113 (0.094 to 0.133) & $-0.486$ to 0.712 & 0.599 (0.565 to 0.633) \\
 & 3D-RBA (\mmtwo) & $-4.4$ ($-5.9$ to $-3.0$) & $-50.4$ to 41.6 & 46.0 (40.9 to 51.2) \\
\bottomrule
\end{tabularx}
\par\smallskip\footnotesize
Bias is Reader A minus Reader B. All estimates use 1,189 examinations from 416 participants. LoA account for participant and residual components of the paired-difference variance. Bias and MDC95 CIs use 2,000 participant-level bootstrap iterations. Overall volume and 3D-RBA are sums across FC, TC, and PC; overall 3D-RT is the unweighted mean of the three regional values. MDC95 estimates measurement error attributable to interreader mask correction within this workflow.
\par\endgroup

\clearpage
\section*{Supplementary figure legends}
\begingroup\linespread{1.0}\selectfont\small
\subhead{Figure S1. Frequency distributions of overall cartilage agreement metrics.}
A--D show Dice, IoU, HD95, and ASD, respectively. Dashed lines indicate medians. Distributions describe central tendency and tails of overall-cartilage agreement between initial AI masks and adjudicated reference masks.

\subhead{Figure S2. Structure-specific agreement at principal visits.}
A--D show Dice, IoU, HD95, and ASD. V0, V6, and V8 include 403, 393, and 392 examinations, respectively. Colors distinguish visits; yellow triangles indicate means. Boxes, whiskers, and focused vertical ranges follow main Figure 1. One nonprincipal examination is excluded.

\subhead{Figure S3. Agreement in the V1.0 deployment subset.}
The subset includes 700 examinations. A--D show Dice, IoU, HD95, and ASD. The figure describes correspondence between initial V1.0 masks and adjudicated references; boxplots, means, and axis ranges follow main Figure 1. Versions were deployed in different imaging subsets; stratified results are descriptive.

\subhead{Figure S4. Agreement in the V2.0 deployment subset.}
The subset includes 489 examinations. A--D show Dice, IoU, HD95, and ASD. The figure describes correspondence between initial V2.0 masks and adjudicated references; boxplots, means, and axis ranges follow main Figure 1. Versions were deployed in different imaging subsets; stratified results are descriptive.

\subhead{Figure S5. Spatial agreement across 120 three-region geometric experiments.}
Each panel shows the distribution and individual values of one area-weighted spatial metric across the 40 FC, 40 TC, and 40 PC experiments. Boxes span the 25th--75th percentiles, green lines indicate medians, yellow triangles indicate means, and pale blue points show individual experiments.

\subhead{Figure S6. Absolute percentage errors by participant and cartilage region.}
All 40 participants and three regions are displayed in the prespecified sample order. Cell labels show the unrounded-data absolute percentage error to one decimal place. The nonlinear color scale spans 0--45\% and retains the 43.1\% FC outlier without clipping.

\subhead{Figure S7. Repeated-measures Bland--Altman agreement for overall cartilage quantification.}
A--C show total cartilage volume, 3D-RT mean thickness, and 3D-RBA defect area, respectively. Each point represents one examination; horizontal coordinates are paired reader means and vertical coordinates are Reader A minus Reader B. Red lines indicate mean biases, dashed lines indicate repeated-measures 95\% LoA, and orange lines show linear trends. All 1,189 examinations from 416 participants were included. Point estimates account for participant and residual components of paired-difference variance; confidence intervals are reported in Table S6.
\par\endgroup

\splate{1}{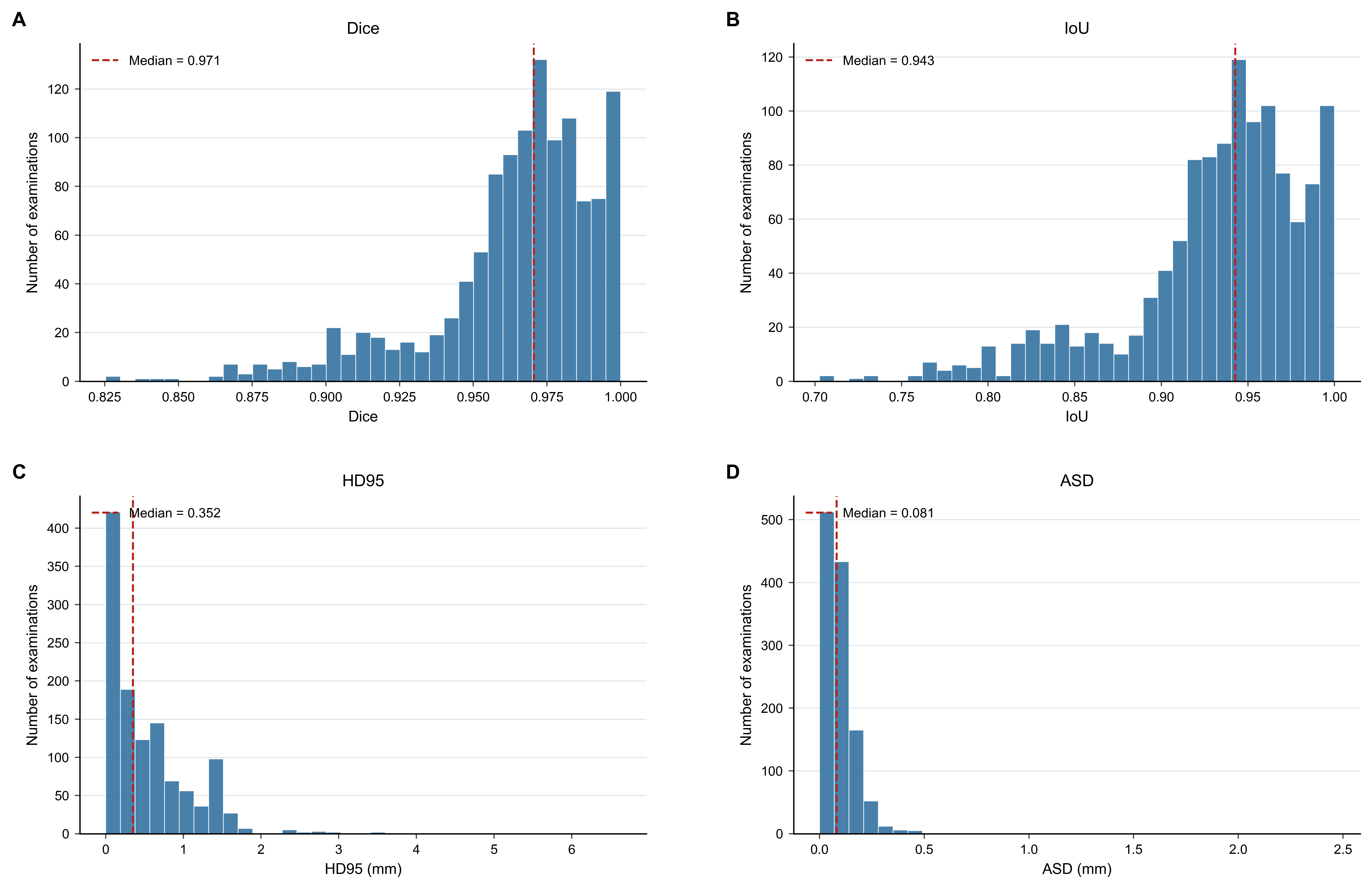}{Overall-cartilage agreement distributions}
\splate{2}{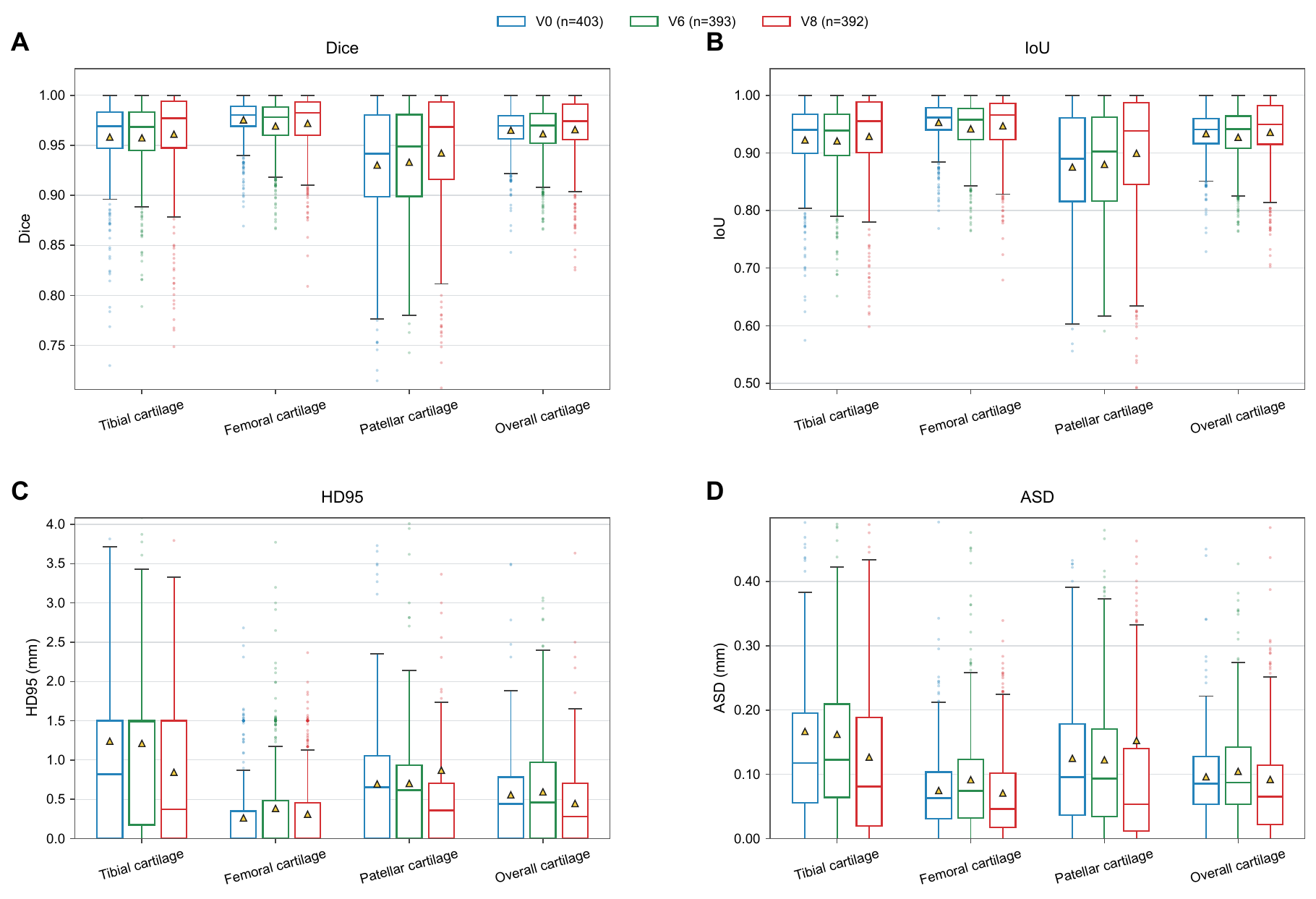}{Agreement by principal visit}
\splate{3}{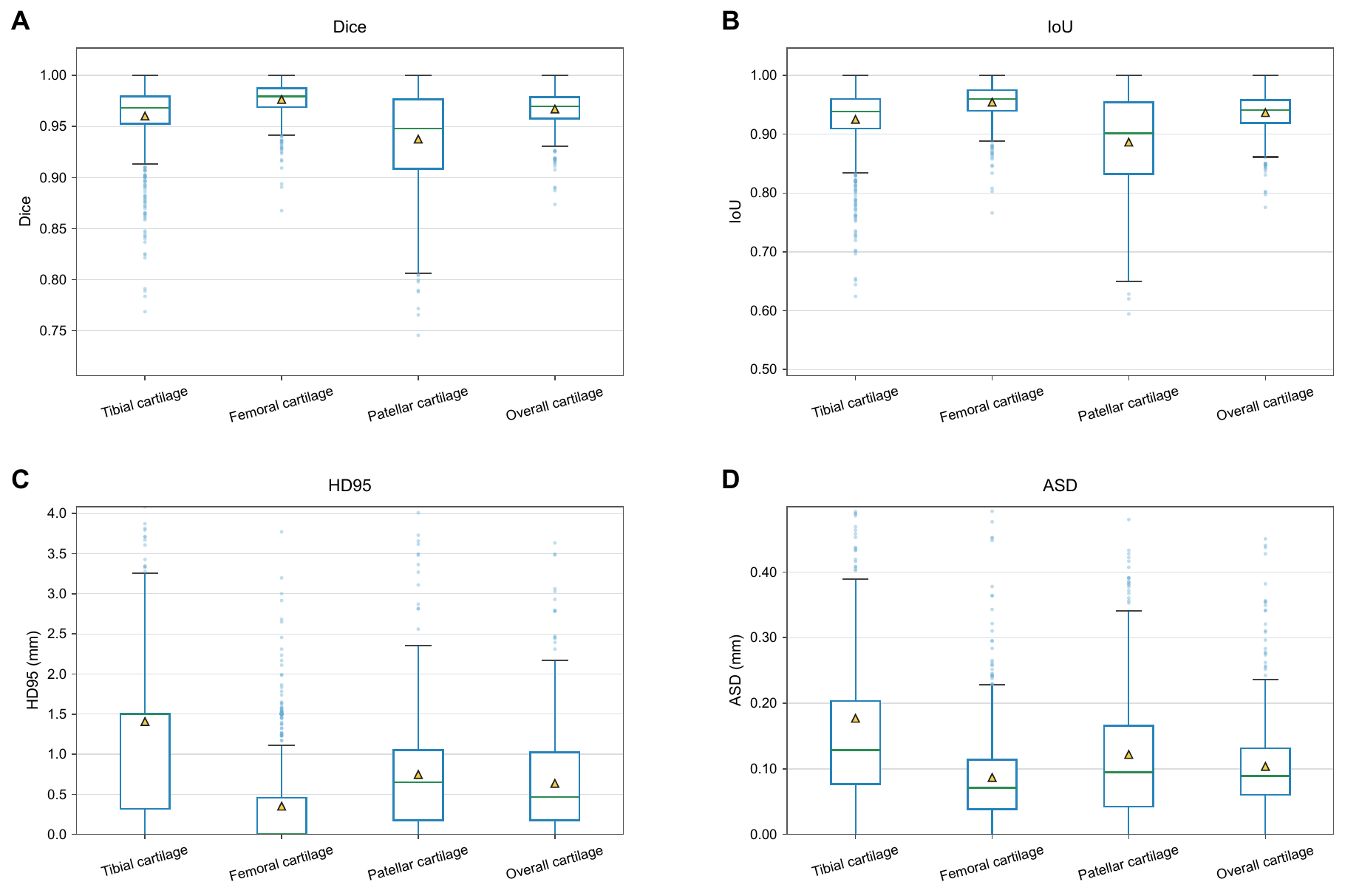}{V1.0 deployment agreement}
\splate{4}{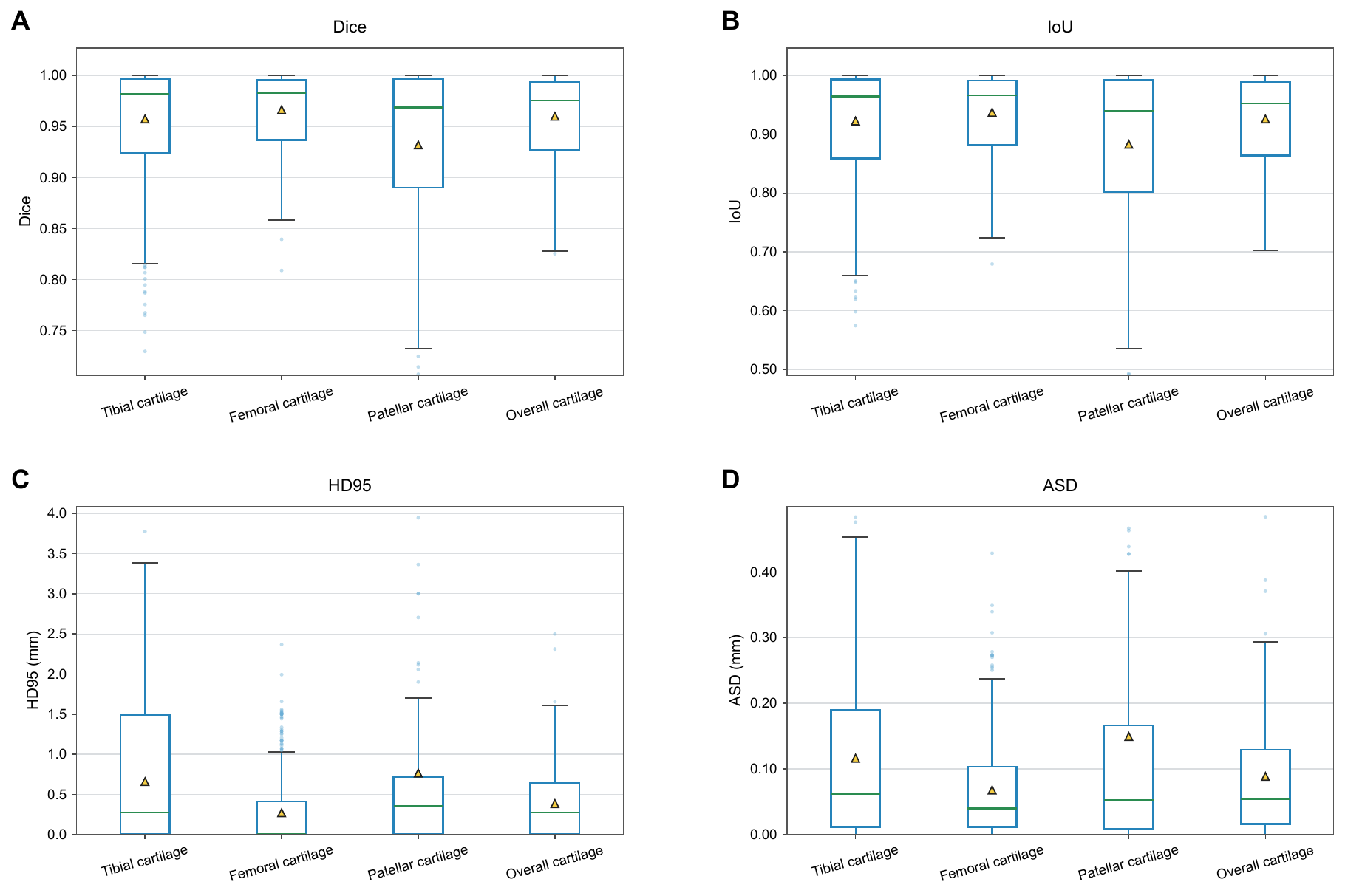}{V2.0 deployment agreement}
\splate{5}{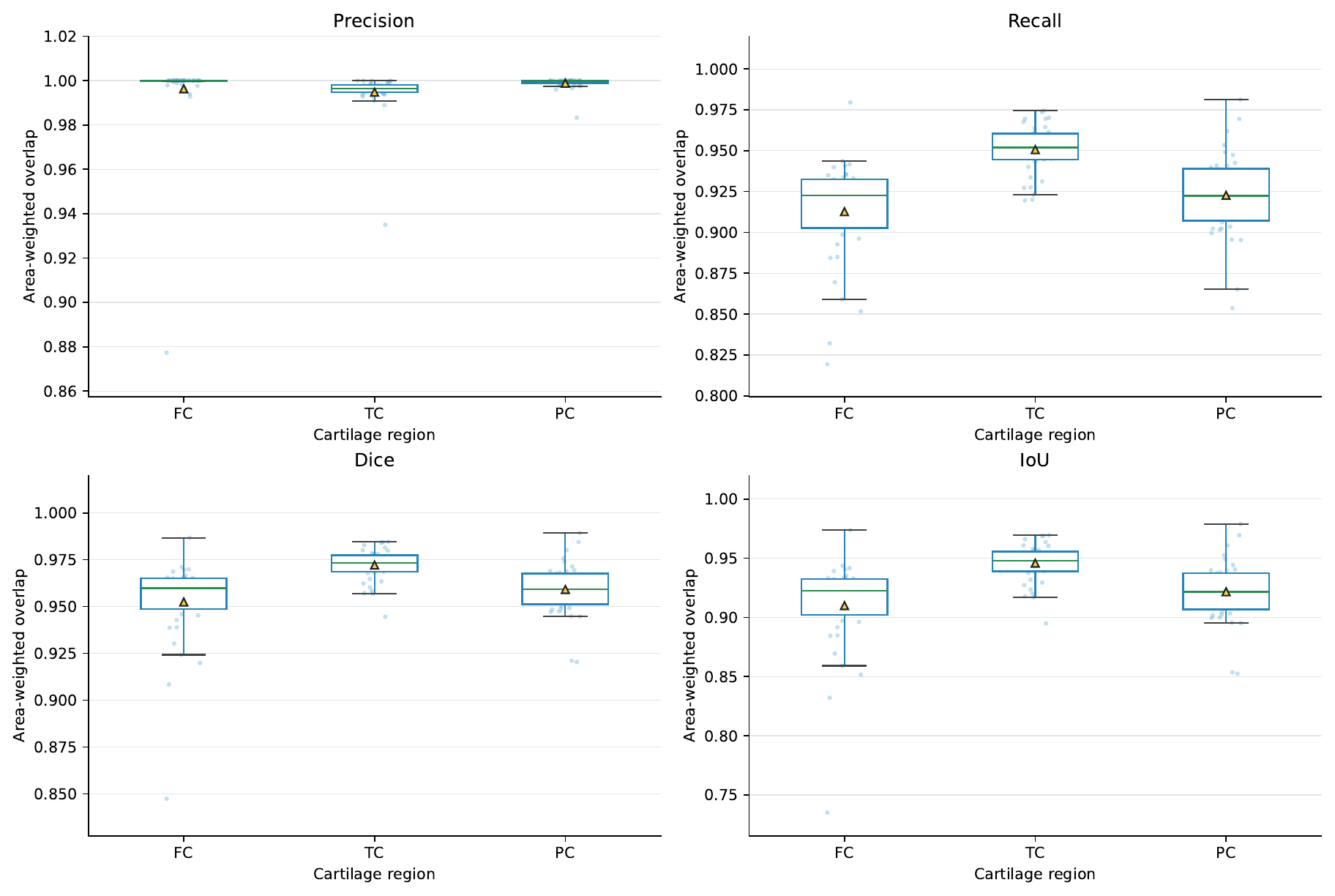}{Experiment-level geometric results}
\splate{6}{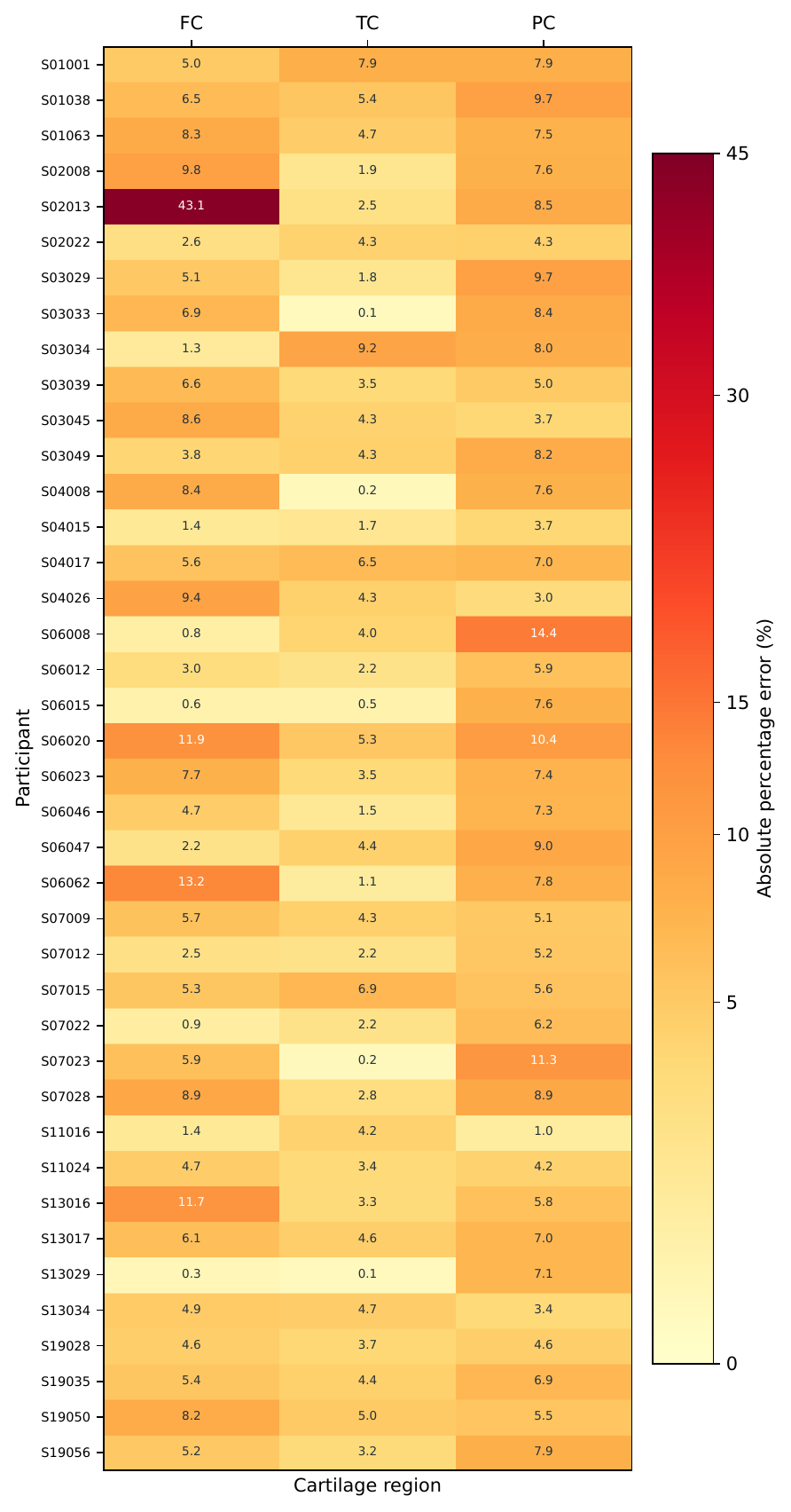}{Absolute percentage error heatmap}
\splate{7}{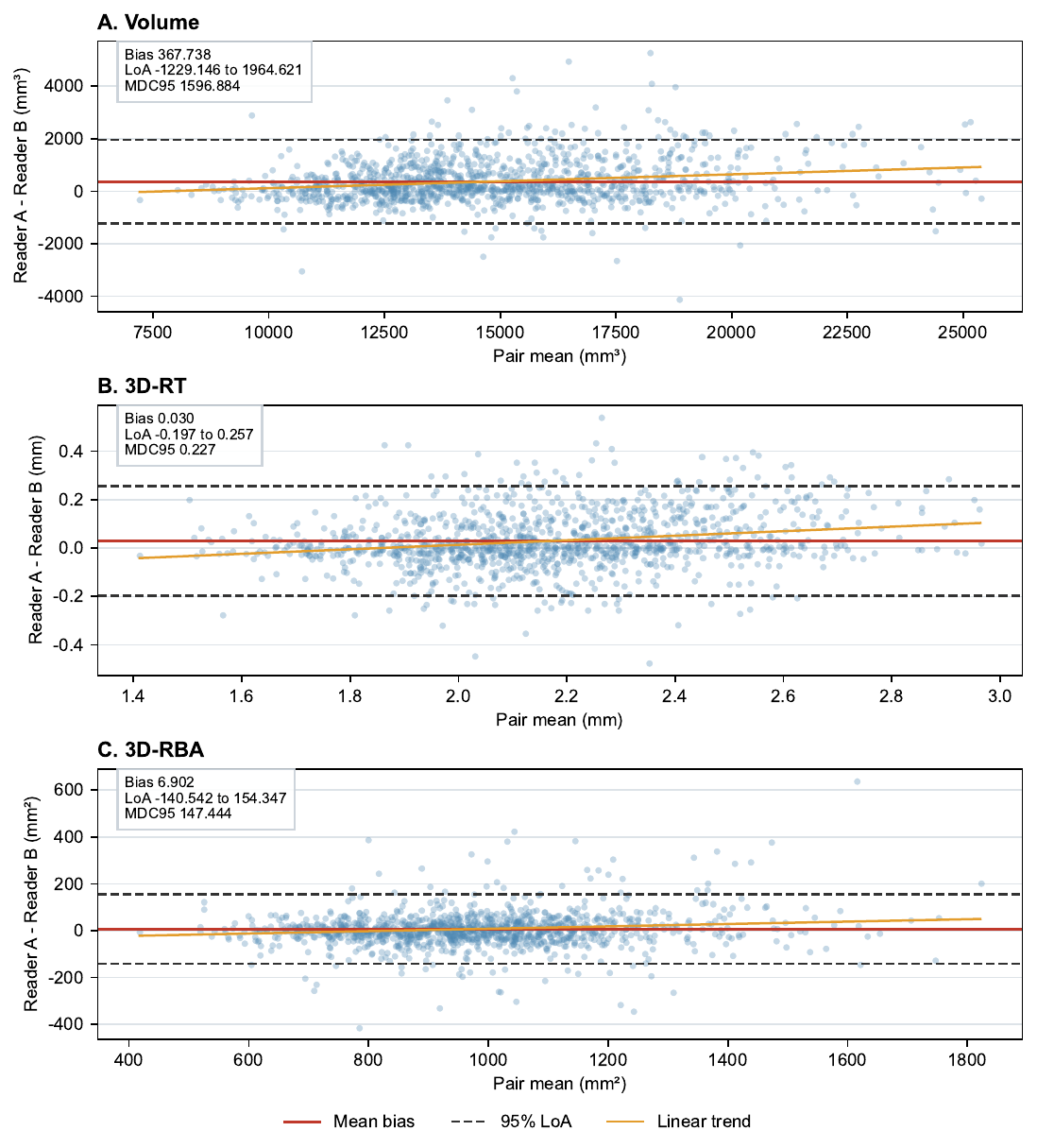}{Repeated-measures Bland--Altman agreement}

\clearpage
\section*{Graphical abstract}
\begin{center}
\includegraphics[width=\linewidth,height=0.65\textheight,keepaspectratio]{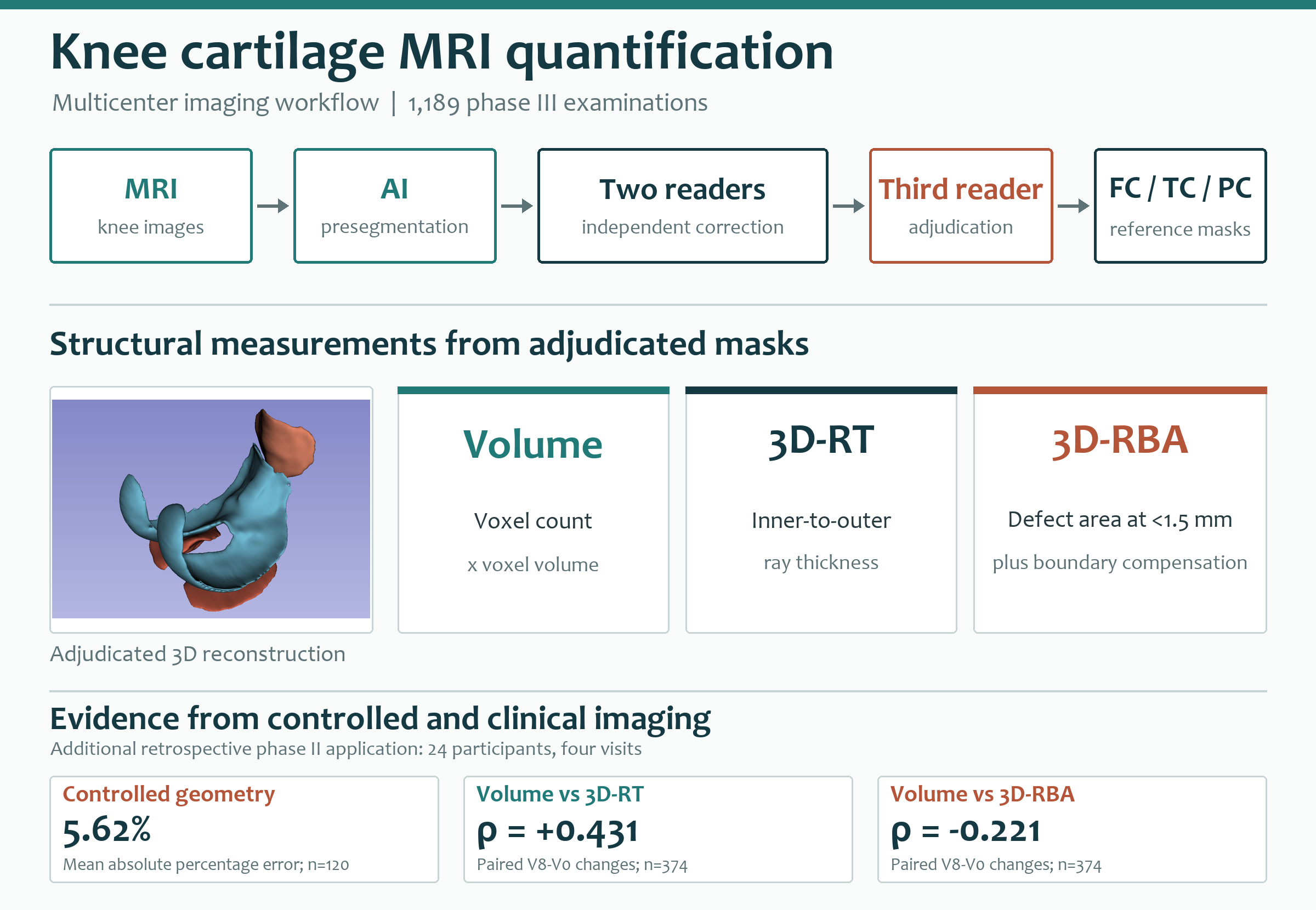}
\end{center}
An AI-assisted, adjudicated MRI workflow quantified knee cartilage volume, ray-traced thickness, and threshold-defined defect area. 3D-RBA showed an interreader intraclass correlation coefficient of 0.932 and 5.62\% mean absolute percentage error in controlled geometry; among 374 phase III participants, volume changes correlated positively with thickness and negatively with defect area.

\end{document}